\documentclass{aa}
\usepackage{graphicx}
\usepackage{natbib}
\usepackage{txfonts}
\pdfoutput=1

\def\Ha{\ifmmode^{\mathrm{H}\alpha }\else$\mathrm{H}\alpha$\fi}
\def\Hb{\ifmmode^{\mathrm{H}\beta }\else$\mathrm{H}\beta$\fi}
\def\LyA{\ifmmode^{\mathrm{H}\alpha }\else$\mathrm{Ly}\alpha$\fi}
\def\BrA{\ifmmode^{\mathrm{Br}\alpha }\else$\mathrm{Br}\alpha$\fi}
\def\BrG{\ifmmode^{\mathrm{Br}\gamma }\else$\mathrm{Br}\gamma$\fi}
\def\PaB{\ifmmode^{\mathrm{Pa}\beta }\else$\mathrm{Pa}\beta$\fi}
\def\mag{\ifmmode^{\rm m }\else$^{\rm m}$\fi}

\def\as{$\,^{\prime\prime}\,$}

\def\hh{\ifmmode^{\rm h}\else$^{\rm h}$\fi}
\def\mm{\ifmmode^{\rm m}\else$^{\rm m}$\fi}
\def\ss{\ifmmode^{\rm s}\else$^{\rm s}$\fi}
\def\deg{\ifmmode^\circ\else$^\circ $\fi}
\def\amin{\ifmmode^\prime\else$^\prime $\fi}

\def\decdm#1#2{\ifmmode{#1}\else{$#1$}\fi\deg\ #2\amin\ }

\def\dec#1#2#3{\ifmmode{#1}\else{$#1$}\fi\deg\ #2\amin\ #3\as\ }

\def\decb#1#2#3#4{\ifmmode{#1}\else{$#1$}\fi\deg\ #2\amin\ #3\farcs#4 }
\def\toria{$\theta^1$Ori\,A}
\def\torib{$\theta^1$Ori\,B}
\def\toric{$\theta^1$Ori\,C}
\def\toricone{$\theta^1$Ori\,C1}
\def\torid{$\theta^1$Ori\,D}
\def\torionisc{$\theta^1$Orionis\,C}

\begin{document}

\title{
  Tracing the young massive high-eccentricity binary system {\torionisc} 
  through periastron passage
  \thanks{Based on observations
    made with ESO telescopes at the La Silla Paranal Observatory under the OT and VISA-MPG GTO programme IDs
    078.C-0360(A), 080.C-0541(A,B,C,D), 080.D-0225(B), and 080.C-0388(A).
  }
}

\titlerunning{Tracing the high-eccentricity binary {\toric} through periastron passage}

\author{
  S.~Kraus\inst{1} \and
  G.~Weigelt\inst{1} \and
  Y.~Y.~Balega\inst{2} \and
  J.~A.~Docobo\inst{3} \and
  K.-H.~Hofmann\inst{1} \and
  T.~Preibisch\inst{4} \and 
  D.~Schertl\inst{1} \and
  V.~S.~Tamazian\inst{3} \and
  T.~Driebe\inst{1} \and
  K.~Ohnaka\inst{1} \and
  R.~Petrov\inst{5} \and
  M.~Sch\"{o}ller\inst{6} \and
  M.~Smith\inst{7}
}

\authorrunning{Kraus~et~al.}

\offprints{skraus@mpifr-bonn.mpg.de}

\institute{
  Max-Planck-Institut f\"ur Radioastronomie, Auf dem H\"ugel 69, 53121 Bonn, Germany
  \and
  Special Astrophysical Observatory, Russian Academy of Sciences, Nizhnij
  Arkhyz, Zelenchuk region, Karachai-Cherkesia, 357147, Russia
  \and
  Astronomical Observatory R.~M.~Aller, University of Santiago de Compostela, Galicia, Spain
  \and
  Universit\"ats-Sternwarte M\"unchen, Scheinerstr.\ 1, 81679 M\"unchen, Germany
  \and
  Laboratoire Universitaire d'Astrophysique de Nice, UMR 6525 Universit\'{e} de Nice/CNRS, Parc Valrose, 06108 Nice Cedex 2, France
  \and
  European Southern Observatory, Karl-Schwarzschild-Str.~2, 85748 Garching, Germany
  \and
  Centre for Astrophysics \& Planetary Science, University of Kent, Canterbury CT2 7NH, UK
}

\date{Received 2008-06-11; accepted 2009-01-27}

\abstract
    {
      The nearby high-mass star binary system {\toric} is the brightest and most massive 
      of the Trapezium OB stars at the core of the Orion Nebula Cluster, 
      and it represents a perfect laboratory to determine the 
      fundamental parameters of young hot stars and to constrain the distance of the 
      Orion Trapezium Cluster.
    }
    {
      By tracing the orbital motion of the {\toric} components, 
      we aim to refine the dynamical orbit of this important binary system.
    }
    {
      Between January 2007 and March 2008, we observed {\toric} with 
      VLTI/AMBER near-infrared ($H$- and $K$-band) long-baseline interferometry, 
      as well as with bispectrum speckle interferometry with the ESO~3.6\,m and the BTA~6\,m
      telescopes ($B'$- and $V'$-band).
      Combining AMBER data taken with three different 3-telescope array configurations, 
      we reconstructed the first VLTI/AMBER closure-phase aperture synthesis image,
      showing the {\toric} system with a resolution of $\sim 2$\,mas.
      To extract the astrometric data from our spectrally dispersed AMBER data,
      we employed a new algorithm, which fits the wavelength-differential visibility
      and closure phase modulations along the $H$- and $K$-band and is insensitive to
      calibration errors induced, for instance, by changing atmospheric conditions.
    }
    {
      Our new astrometric measurements show that the companion has 
      nearly completed one orbital revolution since its discovery in 1997.
      The derived orbital elements imply a short-period ($P \approx 11.3$~yr) 
      and high-eccentricity orbit ($e \approx 0.6$) with periastron passage around 2002.6.
      The new orbit is consistent with recently published radial velocity measurements,
      from which we can also derive the first direct constraints on the mass ratio
      of the binary components.
      We employ various methods to derive the system mass ($M_{\rm system}=44 \pm 7~M_{\sun}$) 
      and the dynamical distance ($d=410 \pm 20$~pc), which is in remarkably good agreement with
      recently published trigonometric parallax measurements obtained with radio interferometry.
    }
    {}
    
    
    \maketitle

%

\section{Introduction}\label{sec:intro}

The Orion Nebula (M42) is one of the closest ($d \sim 400-450$ pc)
and most prominent star-forming regions (see \citealp{ode01} for a review).
It contains a massive cluster of very young
($\sim 1\times 10^6$ yr) stars \citep[cf.][]{her86, mcc94, hil97}, which is
known as the Orion Nebula Cluster (ONC).  
Due to its relatively close distance and
its favorable celestial location (which makes it accessible to
observatories on the northern and southern hemispheres), the
ONC is probably the best investigated young cluster in the whole sky
and has been observed at virtually every wavelength.  
It is a perfect laboratory for observations of
young stellar objects over the full mass range, from very low-mass
brown dwarfs to massive O-type stars.

The brightest star in the cluster is the massive O7--O5.5 type
\footnote{\citet{sim06} determined the stellar effective temperature of
{\toric} to be $T_{\rm eff}=39\,000 \pm 1\,000$~K with $\log g=4.1$~dex.
In recent stellar atmosphere models \citep{mar05}, this corresponds to a spectral type
of O5.5--O6, while the same parameters correspond to a
later spectral type of O7.5--O8 using earlier models \citep[e.g.][]{vac96}.}
star {\toric}, which is known to be a close visual binary system.
After the initial discovery of the companion at
a separation $\rho$ of $0\farcs033$ (33\,mas, corresponding to about 15~AU)
with near-infrared (NIR) bispectrum speckle interferometry
by \citet{wei99},
\citet{sch03a} have presented further observations
and reported the first detection of orbital motion.
\citet{kra07} presented the first speckle observations at visual
wavelengths, the first NIR long-baseline
interferometric observations of {\toric} using the IOTA
interferometer, and produced an aperture-synthesis image of the system.
They also performed a joint analysis of all existing interferometric
measurements that covered a period of more than 9 years and clearly revealed
orbital motion.  After reaching a maximum value of 42\,mas in 1999,
the separation of the system steadily decreased to 13\,mas in 2005.
Detailed modeling of these data yielded a preliminary orbit solution
with a high eccentricity ($e\sim 0.91$) and a period
of 10.9~yrs. According to this solution, the periastron passage should
have occurred around July 2007 with a closest separation of less than 2~AU.
\citet{pat08} recently presented additional interferometric observations 
of {\toric} obtained with NPOI at visual wavelengths.
Extending the orbital coverage by about 1.2~yrs, they measured a
companion position which deviates $\sim 4$\,mas 
from the position predicted by the orbital solution of \citet{kra07}
and concluded that the orbit has a considerably lower eccentricity 
($e \sim 0.16$) and a longer period ($\sim 26$~yrs).
This shows a clear need for further interferometric observations
that will clarify this apparent discrepancy.
A reliable and accurate orbit solution will also provide unique constraints
to the basic stellar parameters and, most importantly, will yield the
masses of the stars and the distance.

The orbital motion of the companion might also be responsible for the
radial velocity variations reported by \citet{sta96,sta08} and others.
Due to the incomplete coverage and the significant scatter in 
the derived velocities, it is not yet possible to derive the orbital
elements of the spectroscopic orbit \citep{sta08}. 

In spite of its importance and the huge number of studies,
the distance to the ONC is, even after decades of investigation,
still not well known and an issue of ongoing discussion.
The ``canonical'' value of 450~pc \citep{her86,gen89} that was 
widely used during the last two decades was recently
challenged by some studies finding significantly smaller values.
Numerous new distance determinations 
\citep[e.g. ][]{sta06, jef07, may08},
also including the first direct trigonometric parallax measurements on radio sources in the ONC by 
  \citet[][ $437 \pm 19$~pc]{hir07}, 
  \citet[][ $389^{+24}_{-21}$~pc]{san07}, and
  \citet[][ $414 \pm 7$~pc]{men07}, 
  yielded distances mainly in the range between 390~pc and 440~pc.
The $\sim 13\%$ difference in these distance estimates,
which to some extent might also include an intrinsic distance spread 
of the studied stars, leads to an
$\sim 30\%$ uncertainty in the derived stellar luminosities and
correspondingly affects any age and mass estimates for the stars.
This is a serious limitation to the usefulness of the exceptionally
well-studied young stellar population; 
e.g., for the calibration of pre-main-sequence evolutionary models
or in the investigation of the spatial relationship between the
young stellar groups in the wider region of the Orion association.
Therefore, trigonometric parallax measurements of stars in the core 
of the Trapezium OB star cluster (e.g.\ on GMR~12={\toria}2, for which \citealt{men07}
measured $418 \pm 9$~pc) are highly desirable.
An alternative way to obtain such distance estimates is through astrometric measurements 
on close binary systems such as {\toric}, yielding the dynamical parallax of the system.

\section{Observations and data reduction}
\label{sec:observations}

\begin{table*}[t]
\caption{Observation log for our new bispectrum speckle and long-baseline interferometric observations.}
\label{tab:observations}
\centering

\begin{tabular}{lllcclcl}
  \hline\hline
  Instrument         & Date        & UT           & Telescope     & Spectral           & DIT                     & No.\ Interferograms & Calibrator(s)\\
                     & [UT]        &              & Triplet       & Mode               &                         & Target/Calibrator  & \\
  \noalign{\smallskip}
  \hline
  \noalign{\smallskip}
  BTA~6\,m/Speckle   & 2007 Nov 25 &              &              & 550\,nm/20\,nm     & 20\,ms                  & 1940/1940          & {\torid}\\
  ESO~3.6\,m/Speckle & 2008 Jan 10 &              &              & 440\,nm/16\,nm     & 10\,ms                  & 18000/10000        & \object{36~Ori}\\
  ESO~3.6\,m/Speckle & 2008 Jan 10 &              &              & 550\,nm/30\,nm     & 10\,ms                  & 10000/6000         & 36~Ori\\
  BTA~6\,m/Speckle   & 2008 Jan 26 &              &              & 550\,nm/20\,nm     & 20\,ms                  & 1500/2000          & {\torid}\\
  \hline
  VLTI/AMBER         & 2007 Jan 08 & 04:36, 06:32 & UT1-UT3-UT4  & LR-$K$               & 26\,ms                  & 12000/10000        & HD\,41547\\
  VLTI/AMBER         & 2007 Dec 03 & 07:19        & A0-D0-H0     & LR-$HK$              & 26\,ms                  & 5000/10000         & HD\,33833\\
  VLTI/AMBER         & 2007 Dec 03 & 07:34, 08:44 & A0-D0-H0     & LR-$HK$              & 50\,ms                  & 7500/7500          & HD\,33833\\
  VLTI/AMBER         & 2007 Dec 03 & 07:48, 08:54 & A0-D0-H0     & LR-$HK$              & 100\,ms                 & 2000/750           & HD\,33833\\
  VLTI/AMBER         & 2007 Dec 05 & 06:05        & A0-K0-G1     & LR-$HK$              & 26\,ms                  & 5000/5000          & HD\,33833\\
  VLTI/AMBER         & 2007 Dec 05 & 05:49, 07:28 & A0-K0-G1     & LR-$HK$              & 50\,ms                  & 10000/15000        & HD\,33833\\
  VLTI/AMBER         & 2007 Dec 05 & 06:17, 07:46 & A0-K0-G1     & LR-$HK$              & 100\,ms                 & 3500/2500          & HD\,33833\\
  VLTI/AMBER         & 2008 Feb 22 & 03:44        & A0-D0-H0     & LR-$HK$              & 50\,ms                  & 5000/10000         & HD\,37128, HD\,50281\\
  VLTI/AMBER         & 2008 Feb 22 & 04:21        & A0-D0-H0     & LR-$HK$              & 100\,ms                 & 2500/3500          & HD\,37128, HD\,50281\\
  VLTI/AMBER         & 2008 Feb 24 & 03:11        & A0-K0-G1     & LR-$HK$              & 50\,ms                  & 10000/5000         & HD\,37128\\
  VLTI/AMBER         & 2008 Feb 24 & 02:37        & A0-K0-G1     & LR-$HK$              & 100\,ms                 & 6000/4000          & HD\,37128\\
  VLTI/AMBER         & 2008 Mar 03 & 02:31        & D0-H0-G1     & LR-$HK$              & 50\,ms                  & 5000/5000          & HD\,43023\\
  VLTI/AMBER         & 2008 Mar 03 & 02:52        & D0-H0-G1     & LR-$HK$              & 100\,ms                 & 1500/1500          & HD\,43023\\
  \noalign{\smallskip}
  \hline
\end{tabular}
\end{table*}

\begin{table}[t]
\caption{AMBER calibrator stars and their characteristics, including uniform disk (UD) diameters. }
\label{tab:calibrators}
\centering
\begin{tabular}{lccccc}
  \hline\hline
  Star                & $V$ & $H$ & $K$ & Spectral & Adopted UD diameter\\
                      &   &   &   & Type     & [mas]\\
  \noalign{\smallskip}
  \hline
  \noalign{\smallskip}
  \object{HD\,33833}  & 5.9 & 3.9 & 3.8 & G7III & $0.83 \pm 0.06$$^{a}$\\
  \object{HD\,37128}  & 1.7 & 2.4 & 2.3 & B0I   & $0.86 \pm 0.16$$^{b}$\\
  \object{HD\,41547}  & 5.9 & 5.1 & 5.0 & F4V   & $0.41 \pm 0.03$$^{a}$\\
  \object{HD\,43023}  & 5.8 & 3.7 & 3.5 & G8III & $0.98 \pm 0.07$$^{a}$\\
  \object{HD\,50281}  & 6.6 & 4.3 & 4.1 & K3V   & $0.77 \pm 0.10$$^{c}$\\
  \noalign{\smallskip}
  \hline
\end{tabular}

\begin{flushleft}
  {\it Notes}~--~$^{a}$ UD diameter computed with ASPRO \\
  \hspace{12mm}{\tt (http://www.jmmc.fr/aspro\_page.htm)}.\\
  \hspace{10.5mm}$^{b}$ UD diameter taken from \citet{moz91}.\\
  \hspace{10.5mm}$^{c}$ UD diameter taken from getCal tool \\
  \hspace{12mm}{\tt (http://mscweb.ipac.caltech.edu/gcWeb/)}.\\
\end{flushleft}
\end{table}

\subsection{Bispectrum speckle interferometry}
\label{sec:obsspeckle}

\begin{figure*}[t]
  \centering
  \includegraphics[width=18cm]{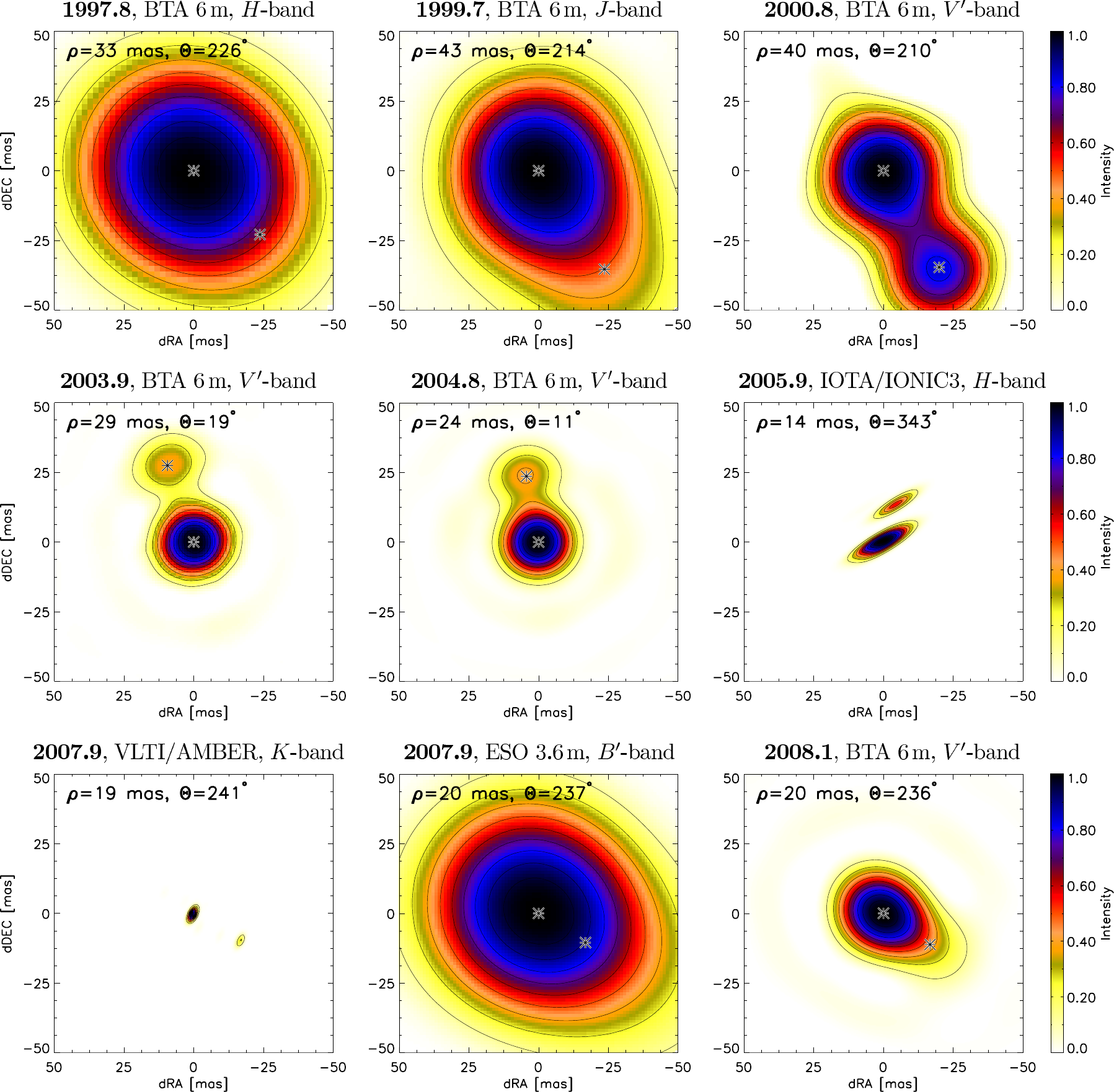}\\
  \caption{
    Selection of interferometric images of the {\toric} system obtained by our group between 1997 and 2008,
    revealing the orbital motion of the companion.
    The images were reconstructed either from $H$-, $J$-, $V'$-, or $B'$-band speckle 
    interferograms recorded at the BTA 6~m or the ESO 3.6\,m telescopes, or from
    $H$-band Michelson interferograms recorded with the IOTA 3-telescope interferometer
    (see \citealt{kra07} for details).  
    For epoch 2007.9, we show the aperture synthesis image which we reconstructed from
    VLTI/AMBER data using the procedure described in Sect.~\ref{sec:imaging}.
    For each image, 10\% intensity level contours are shown.
    In the speckle images, the fitted component positions are marked with an asterisk.
  }
  \label{fig:image}
\end{figure*}

Bispectrum speckle interferometry is a powerful technique to overcome atmospheric
perturbations and to reach the diffraction-limited resolution of ground-based
telescopes at visual wavelengths.
After monitoring the orbital evolution of the system between 1997 and 2004 at wavelengths 
between 2.1~$\mu$m and 545~nm \citep[$\rho=43\ldots24$\,mas; ][]{wei99,sch03a,kra07} using the Russian 
\textit{Big Telescope Alt-azimuthal} (BTA) 6.0~m telescope, {\toric} became unresolvable with
6~m-class telescopes in 2006.
In November 2007 ($\rho=20$\,mas) and January 2008, the system could again be resolved with the BTA~6\,m 
telescope using a $V'$ medium-band filters ($\lambda_{c}=550$~nm, $\Delta\lambda=20$~nm).   
For these observations, a 1280$\times$1024 
pixel CCD with a multi-alkali S25 intensifier photocathode was used.

In January 2008, additional speckle measurements using $V'$ ($\lambda_{c}=550$~nm, $\Delta\lambda=30$~nm)
and $B'$ ($\lambda_{c}=440$~nm, $\Delta\lambda=16$~nm) medium-band filter were obtained
with the ESO~3.6\,m telescope. 
The speckle observations were carried out in the course of ESO open-time programme 
080.C-0388(A) using our visitor speckle camera, which employs a Marconi/EEV 
electron-multiplying CCD.
For the speckle observations, we recorded interferograms of {\toric} and of nearby unresolved stars
in order to compensate for the atmospheric speckle transfer function.  The calibrator stars,
the number of recorded interferograms and the detector integration times (DITs) are listed in
Table~\ref{tab:observations}.
The modulus of the Fourier transform of the object (visibility) was obtained
with the speckle interferometry method~\citep{lab70}.  For image
reconstruction we used the bispectrum speckle interferometry method
(\citealt{wei77}, \citealt{wei83}, \citealt{loh83}, \citealt{hof86}).

Besides providing an independent astrometric measurement, 
our bispectrum speckle interferometric measurement is of special importance  
as it allows us to solve the 180\degr-ambiguity, which is inherent to
long-baseline interferometric investigations which do not include the closure 
phase \citep[e.g.][]{pat08} or for which the instrumental closure phase sign 
has not yet been calibrated (as for VLTI/AMBER).
Solving this ambiguity is essential for deriving the orbit of the system.
Therefore, we paid special attention while deriving the field 
orientation of our speckle images, using calibration measurements which were taken
with the same instrument setup during the same night as the {\toric} observations
on \object{\toria}, \object{\torib} and the well-studied multi-component object \object{$\eta$~Carinae}.
For the January 2008 observations,
a position angle calibration with an accuracy of $\sim 0.3^{\circ}$ was done using
$K'$-band observations covering the Trapezium stars $\theta^1$Ori\,A, B, and E, followed
by the $V'$ and $B'$ observations of {\torib}, which is a binary system with a separation of $\sim$1\arcsec, where
the fainter component is a close binary with a separtion of $\sim$0\farcs15.
This allows us to unambiguously determine that in January 2008, the fainter component (C2) was
located to the southwest of the primary star (Fig.~\ref{fig:image} and \ref{fig:speckle}). 
Performing a detailed re-analysis on all Speckle data taken by our
group between 1997 and 2005, we found a 180\degr-calibration problem which affected the
position angle of the Speckle measurements at epochs 2003.8, 2003.9254, 2003.928, and 2004.8216,
as published in \citealt{kra07}.
Since these Speckle measurements were used for the calibration of the 
IOTA and NPOI long-baseline interferometric observations by \citet{kra07} 
and \citet{pat08}, a revision of the preliminary orbital solutions 
presented in these papers is required (as presented in Sect.~\ref{sec:orbitsolution}).
In Tab.~\ref{tab:astrometry}, we list all available astrometric data, 
taking the quadrant correction into account.

\subsection{VLTI/AMBER spectro-interferometry}
\label{sec:obsamber}

\begin{figure}[tbp]
  \centering
  \includegraphics[width=8.5cm]{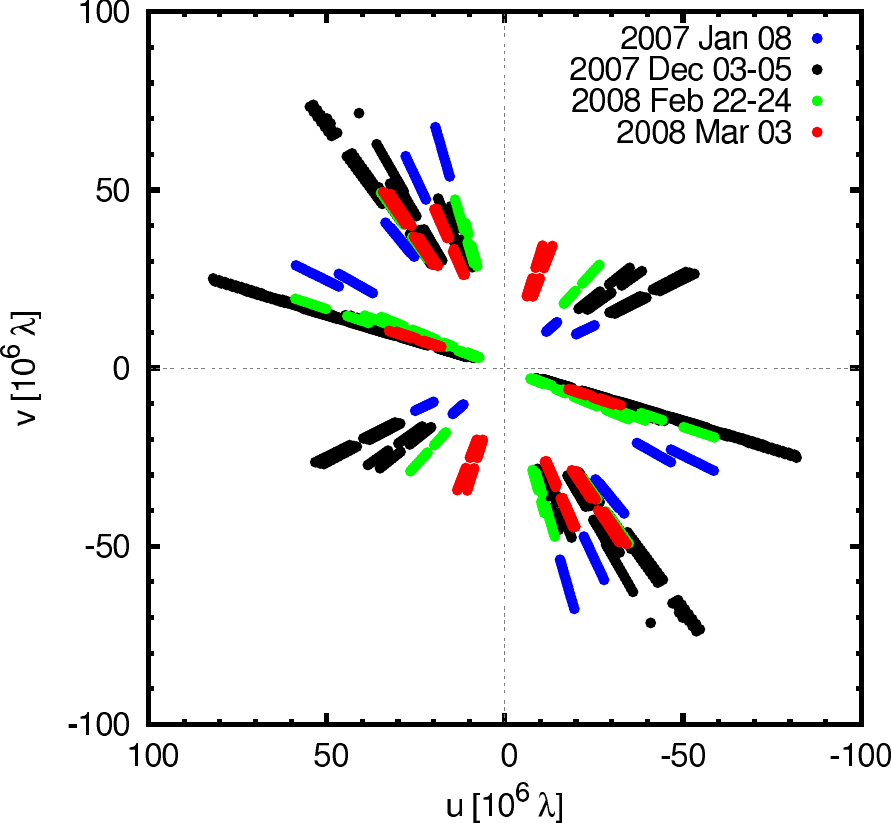}\\
  \caption{$uv$-plane coverage obtained at four 
    epochs on {\toric} with VLTI/AMBER\@.
    The radial extension of the $uv$-tracks reflects the
    spectral coverage of our AMBER interferometric data,
    covering either the $K$-band (LR-$K$ mode, 2007~Jan~08) 
    or the $H$- and $K$-band (LR-$HK$ mode, 2007~Dec~03-05, 2008~Feb~22-24, 2008~Mar~03).
  }
  \label{fig:uvcov}
\end{figure}

AMBER is the NIR interferometric beam-combiner instrument \citep{pet07} of the 
Very Large Telescope Interferometer (VLTI), which is located on Cerro Paranal/Chile and
operated by ESO\@.
For the beam combination, either three 8.2\,m unit telescopes (UTs) or three of the allocatable 
1.8\,m auxiliary telescopes (ATs) can be used.
One outstanding feature of AMBER is its spectral capability, allowing one to observe
several spectral bands with low spectral resolution simultaneously (LR-$JHK$ mode, 
$R=\lambda/\Delta\lambda=35$, covering the $J$-band 
around 1.2~$\mu$m, the $H$-band around 1.6~$\mu$m, and the $K$-band around 2.2~$\mu$m).
The wide wavelength coverage offered by this spectral mode allows us to fit the binary
separation vector with a high accuracy and observing efficiency (as discussed in Sect.~\ref{sec:modelfit}).

The VLTI/AMBER data was recorded in the course of three ESO programmes, 
yielding the {\it uv}-plane coverage shown in Fig.~\ref{fig:uvcov}.
For each science observation, we also recorded interferograms for at least 
one interferometric calibrator star (see Tab.~\ref{tab:calibrators} for the characteristics 
of our calibrator stars), which allows us to calibrate the 
visibilities and the closure phases for instrumental as well as atmospheric effects.

For AMBER data reduction, we employed the {\it amdlib2}-data reduction software
\footnote{The {\it amdlib2} software package is available from the website\\ {\tt 
    http://www.jmmc.fr/data\_processing\_amber.htm}} (release 2.1), which is based on 
the P2VM algorithm~\citep{tat07b}.
For the LR-$HK$ data, we applied the data selection criteria outlined in \citet{kra08}
in order to remove frames which were either degraded by atmospheric effects or
were recorded significantly offset from zero optical path delay.

For the accuracy achievable with our binary model fits, 
the absolute calibration of the wavelength scale is of special importance.
Therefore, we carefully refined the wavelength calibration using the 
telluric gaps between the observed spectral bands.
Using this procedure (which is described in Appendix~\ref{app:speccalib}), 
we reach a calibration accuracy of about $0.03~\mu$m, 
which is still the dominating limiting factor on the total achievable astrometric 
accuracy ($\sim 2$\%).

While the target/calibrator observations from January 2007 were taken 
under good and stable atmospheric conditions, some of the observations in December 2007
and February 2008 suffer from strongly variable seeing conditions
and short atmospheric coherence times.
It is known that short coherence times
can result in a decrease of the measured fringe contrast, which 
might not be completely calibrated out using calibrator measurements.
This effect can impose errors on the absolute calibration, which are expected to
increase with longer DIT and towards shorter wavelengths.
To illustrate this effect, in Fig.~\ref{fig:projectvis} we plot the calibrated 
visibilities measured during our December 2007 observation campaign and compare them to the
cosine visibility modulation of a binary source.
In particular, in the $H$-band the resulting calibration errors can be 
on the order of 20\%.

However, since all spectral channels of an AMBER interferogram are recorded 
at the same time, the wavelength-differential observables (in particular, the 
differential visibility $\delta V(\lambda)$)  are practically insensitive to this 
degradation.
Therefore, in Sect.~\ref{sec:modelfit} we employ a fitting algorithm in which we 
fit only differential visibilities and closure phases in order to determine the 
{\toric} binary separation vector.

A fundamental problem one encounters when interpreting VLTI/AMBER 3-telescope
data concerns the lack of a calibration measurement for the closure phase sign, 
resulting in a 180\degr\ uncertainty on the position angle of binary star observations.
Since our {\toric} VLTI/AMBER observations from December 2007 and February 2008 bracket the
bispectrum speckle measurement from January 2008, our data set allows us 
to unambigously define the closure phase sign for these observations, providing
a direct calibration of the 180\degr\ uncertainty for VLTI/AMBER for the first time.

\section{Modeling} \label{sec:modelfit}

\begin{figure*}[tbp]
  \centering
  \includegraphics[width=18cm]{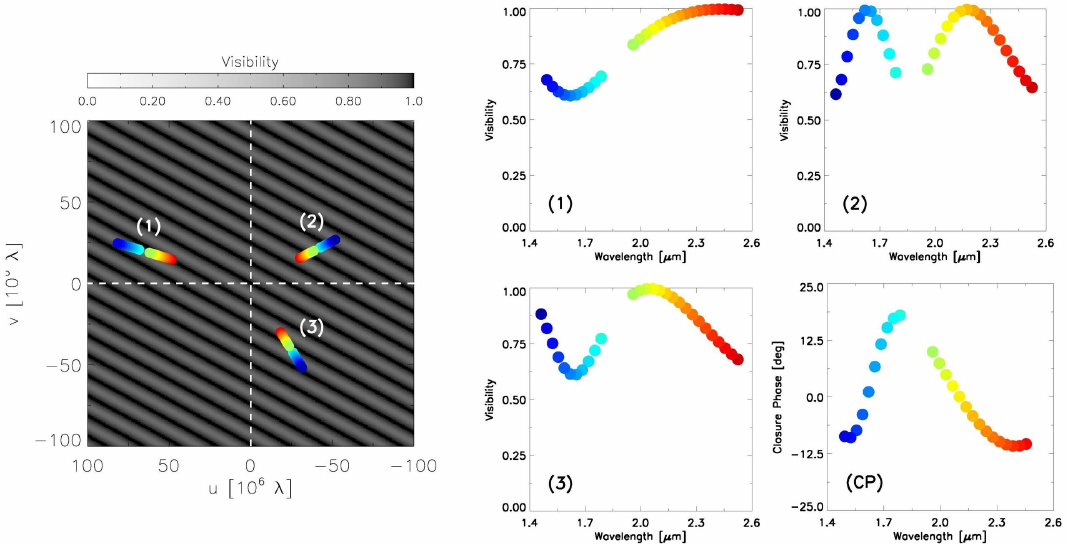}\\
  \caption{Illustration of the basic principle of fitting binary parameters using wavelength-differential interferometric observables. 
    {\it Left:} The VLTI/AMBER 3-telescope interferometer measures the interferometric observables 
    in various spectral channels from $1.4$ to $2.4~\mu$m (as indicated by the color of the dots) and towards three different position angles,
    probing different regions in the two-dimensional Fourier spectrum of the source brightness distribution.
    The figure shows the visibility spectrum for a binary with $\rho=19.07$\,mas and {$\Theta=241.2$\deg} (as inferred for {\toric} on 2007~Dec~03) and the 
    $uv$-sampling obtained with one of our AMBER observations on the same date.
    {\it Middle and Right:} The visibilities and phases show a
    wavelength-differential modulation which is independent of the absolute calibration (see Sect.~\ref{sec:modelfit}) and which can be fitted to analytical models.
  }
  \label{fig:binaryillu}
\end{figure*}

\begin{figure}[htbp]
  \centering
  \includegraphics[width=8.5cm]{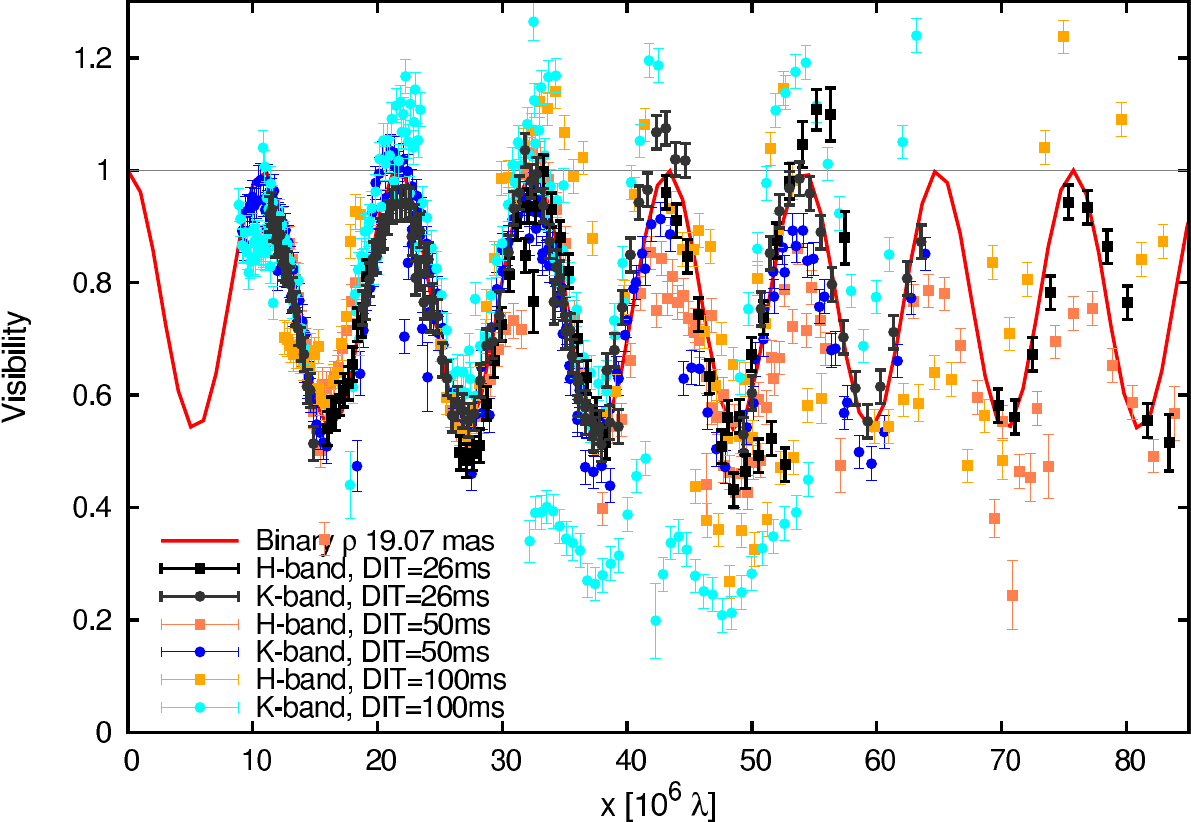}\\
   \caption{AMBER visibilities measured on 2007~Dec~03 and 2007~Dec~05 plotted versus 
    projected distance in the $uv$-plane (where the projection was perpendicular 
    to the fitted binary separation vector $\Theta=241.2\degr$, 
    i.e.\ $x=u\cos(\Theta+90\degr) - v\sin(\Theta+90\degr)$).
    The solid red line shows the theoretical cosine visibility
    profile for a binary star with separation 19.07\,mas and intensity ratio
    0.30.
    As indicated by the strong visibility offsets, which particularly occur at 
    high spatial frequencies, the absolute calibration is sometimes rather poor, 
    reflecting the changing atmospheric conditions during these nights.
    It can also be seen that this calibration bias is particularly important
    for long DITs (50~ms, 100~ms), while it is nearly negligible for short DITs 
    (26~ms, grey \& black points).
    As expected, the spectral dependence of the visibility is not affected 
    by these calibration uncertainties.
  }
  \label{fig:projectvis}
\end{figure}

\begin{figure*}[tbp]
  \centering
  \includegraphics[width=18cm]{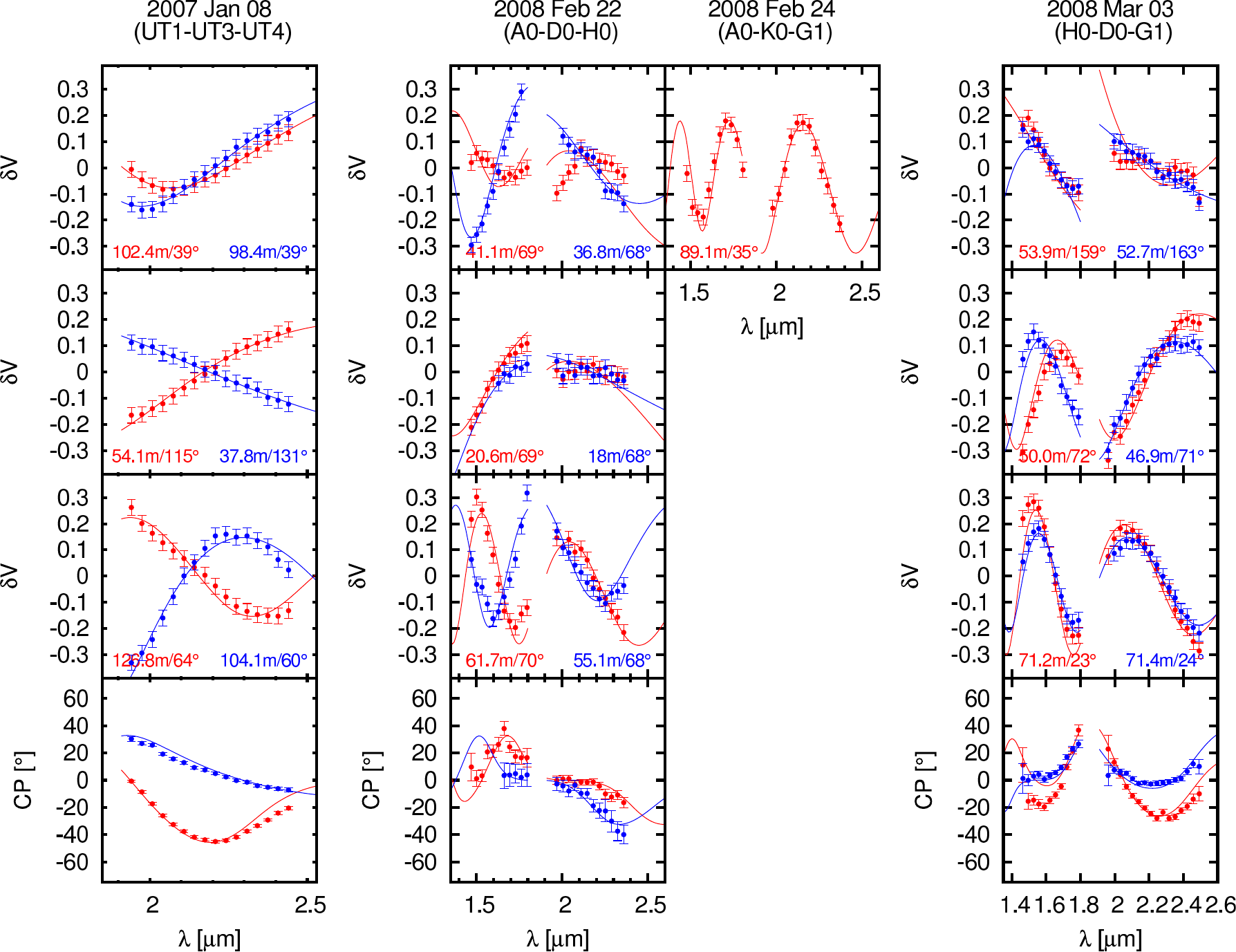}\\
  \caption{Differential visibilities and closure phases measured with VLTI/AMBER in January~2007
    {\it (left)}, February~2008 {\it (middle)}, and March~2008 {\it (right)} on {\toric}.
    The solid lines show the best-fit model corresponding to the binary parameters 
    given in Tab.~\ref{tab:astrometry}.
  }
  \label{fig:fitP78OT}
\end{figure*}

\begin{figure*}[tbp]
  \centering
  \includegraphics[height=14cm]{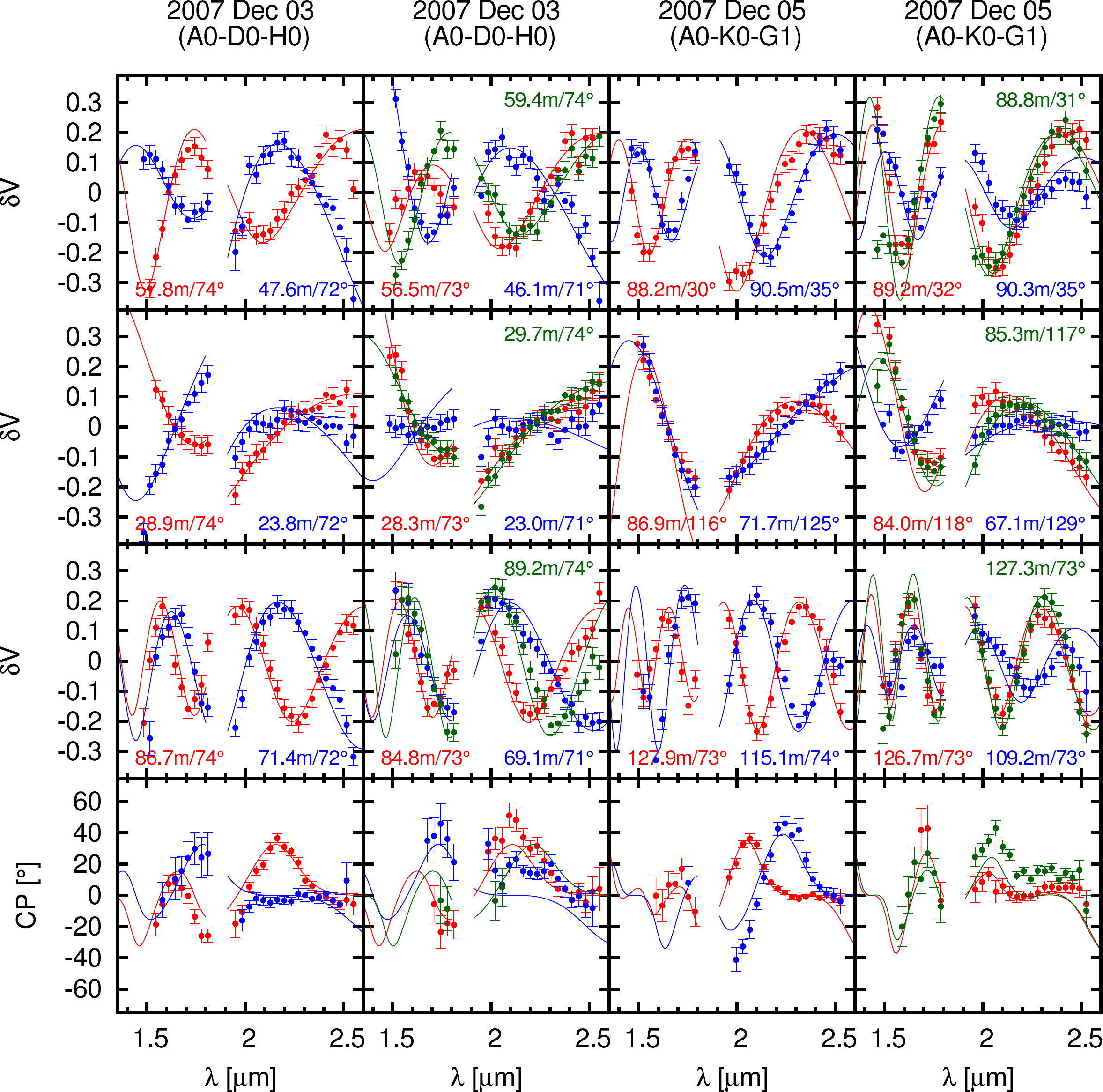}\\
  \caption{Similar to Fig.~\ref{fig:fitP78OT}, showing the AMBER data and
    best-fit model for our observations from December 2007}.
  \label{fig:fitP80Dec}
\end{figure*}

\begin{table*}[t]
\caption{Astrometric measurements for the {\toric} binary system.}
\label{tab:astrometry}
\centering
\begin{tabular}{llccllc}
  \hline\hline
  Telescope           & Date        & Filter & Flux ratio            & \multicolumn{1}{c}{$\Theta$}      & \multicolumn{1}{c}{$\rho$}         & Ref.\\
                      &             &        & $F_{\rm C2}/F_{\rm C1}$ & \multicolumn{1}{c}{[\degr]}       & \multicolumn{1}{c}{[mas]}          & \\
  \noalign{\smallskip} 
  \hline
  \noalign{\smallskip}
  BTA~6\,m/Speckle    & 1997.784    & $H$    & $0.26 \pm 0.02$ & $226.0 \pm 3$   & $33 \pm 2$ & a\\
  BTA~6\,m/Speckle    & 1998.838    & $K'$   & $0.32 \pm 0.03$ & $222.0 \pm 5$   & $37 \pm 4$ & a\\
  BTA~6\,m/Speckle    & 1999.737    & $J$    & $0.31 \pm 0.02$ & $214.0 \pm 2$   & $43 \pm 1$ & b\\
  BTA~6\,m/Speckle    & 1999.8189   & $G'$   & $0.35 \pm 0.04$ & $213.5 \pm 2$   & $42 \pm 1$ & c\\
  BTA~6\,m/Speckle    & 2000.8734   & $V'$   & $0.35 \pm 0.03$ & $210.0 \pm 2$   & $40 \pm 1$ & c\\
  BTA~6\,m/Speckle    & 2001.184    & $J$    & $0.29 \pm 0.02$ & $208.0 \pm 2$   & $38 \pm 1$ & b\\
  BTA~6\,m/Speckle    & 2003.8      & $J$    & $0.30 \pm 0.02$ &  $19.3 \pm 2$   & $29 \pm 2$ & c\\
  BTA~6\,m/Speckle    & 2003.9254   & $V'$   & --              &  $19.0 \pm 2$   & $29 \pm 2$ & c\\
  BTA~6\,m/Speckle    & 2003.928    & $V'$   & --              &  $19.1 \pm 2$   & $29 \pm 2$ & c\\
  BTA~6\,m/Speckle    & 2004.8216   & $V'$   & $0.34 \pm 0.04$ &  $10.5 \pm 4$   & $24 \pm 4$ & c\\
  IOTA                & 2005.92055  & $H$    & $0.28 \pm 0.03$ & $342.74 \pm 2$  & $13.55 \pm 0.5$ & c\\
  NPOI                & 2006.1486   & $V$    & --              & $332.3 \pm 3.5$ & $11.80 \pm 1.11$ & d\\
  VLTI/AMBER          & 2007.0192   & $K$    & $0.31 \pm 0.03$ & $274.9 \pm 1$    & $11.04 \pm 0.5$ & --\\
  NPOI                & 2007.1425   & $V$    & --              & $268.1 \pm 5.2$  & $11.94 \pm 0.31$ & d\\
  NPOI                & 2007.1507   & $V$    & --              & $272.9 \pm 8.8$  & $12.13 \pm 1.58$ & d\\
  NPOI                & 2007.1753   & $V$    & --              & $266.6 \pm 2.1$  & $12.17 \pm 0.37$ & d\\
  NPOI                & 2007.2055   & $V$    & --              & $265.6 \pm 1.9$  & $12.28 \pm 0.41$ & d\\
  NPOI                & 2007.2137   & $V$    & --              & $263.0 \pm 2.3$  & $12.14 \pm 0.43$ & d\\
  BTA~6\,m/Speckle    & 2007.9014   & $V'$        & $0.29 \pm 0.02$      & $238.0 \pm 2$    & $19.8 \pm 2$  &  --\\
  VLTI/AMBER          & 2007.9233   & $H$+$K$    & $0.24 \pm 0.07$       & $241.2 \pm 1$    & $19.07 \pm 0.5$ & --\\
  ESO~3.6\,m/Speckle  & 2008.0274   & $B'$        & $0.29 \pm 0.02$      & $237.0 \pm 3$    & $19.7  \pm 3$  & --\\
  ESO~3.6\,m/Speckle  & 2008.0274   & $V'$        & $0.31 \pm 0.02$      & $236.5 \pm 3$    & $19.6  \pm 3$  & --\\
  BTA~6\,m/Speckle    & 2008.0712   & $V'$        & $0.31 \pm 0.02$      & $236.2 \pm 2$    & $20.1 \pm 2$   & --\\
  VLTI/AMBER          & 2008.1479   & $H$+$K$    & $0.23 \pm 0.09$       & $234.6 \pm 1$    & $21.17 \pm 0.5$ & --\\
  VLTI/AMBER          & 2008.1726   & $H$+$K$    & $0.26 \pm 0.06$       & $236.4 \pm 1$    & $21.27 \pm 0.5$ & --\\
  \noalign{\smallskip}
  \hline
\end{tabular}
\begin{flushleft}
  {\it Notes}~--~The position angles given in col.\ 5 are measured East of North and were corrected for the 180\degr-calibration problem described in Sect.~\ref{sec:obsspeckle}.\\
  {\it References}~--~$(a)$~\citealt{wei99}, $(b)$~\citealt{sch03a}, $(c)$~\citealt{kra07}, $(d)$~\citealt{pat08}.\\
\end{flushleft}
\end{table*}

In order to derive the binary separation $\rho$ and position 
angle~\footnote{Following convention, we measure the position angle (PA) East of North.}
$\Theta$ from the speckle data,
we used the same algorithm as in our earlier studies on this system \citep{wei99,sch03a,kra07}, 
fitting a cosine function directly to the 2-D speckle visibilities. 
For illustration, in the Appendix (Fig.~\ref{fig:speckle}) we show power spectra and
  Fourier spectra determined from three independent data sets obtained during our 
  observing run with the ESO~3.6\,m telescope and compare them with the model power and Fourier spectra
  corresponding to a binary star. 

For the modeling of the AMBER data, we employed an optimized algorithm which fits closure phases (CP)
and differential visibilities $\delta V$.
Each AMBER LR-$HK$ measurement records 16 $K$-band plus 11 $H$-band spectral channels in
the wavelength range from 1.51 to 2.55~$\mu$m, covering about $40$\% of the 
object Fourier spectrum in radial direction (see $uv$-plane tracks in Fig.~\ref{fig:uvcov}).
Strongly resolved objects (such as binaries with a separation of $\rho \gtrsim \Delta\lambda/B$, 
where $\Delta\lambda$ is the recorded spectral window and $B'$ is the projected baseline length) 
can already show significant visibility modulation over this range of spatial frequencies 
(see illustration in Fig.~\ref{fig:binaryillu}).
For instance, our {\toric} AMBER measurements from December 2007
probe spatial frequencies out to the eighth lobe of the cosine binary visibility modulation
(Fig.~\ref{fig:projectvis}) and the wavelength-differential visibilities recorded in 
a single AMBER $H$-/$K$-band measurement sample up to 3 visibility cycles.
This wavelength-differential visibility modulation already provides all information required to determine
the binary separation and orientation.  In this way, the absolute 
calibration of the visibility, which is subject to many adverse atmospheric effects, becomes dispensable.

To determine the differential visibilities $\delta V$ as used in our fitting algorithm,
we first follow the standard data reduction and calibration procedure in order to correct for
wavelength-dependent instrumental effects (using a calibrator measurement taken during the same night).
Then, we remove the absolute calibration by subtracting the average visibility of
the considered spectral window
\begin{equation}
  \delta V(\lambda) = V(\lambda) - \langle V(\lambda) \rangle_{\rm band},
\end{equation}
yielding the differential visibility $\delta V$, which we compute separately for each spectral band
($H$ and $K$-band).

A similar approach is applied to the model visibilities $V_{\mathrm{model}}(\lambda)$ (which 
we compute using equations 7, 11, and 12 from \citealt{kra05}):
\begin{equation}
  \delta V_{\mathrm{model}}(\lambda) = a \left(V_{\mathrm{model}}(\lambda) - b \right),
\end{equation}
where $a$ and $b$ are adjusted to minimize the residuals between the model visibilities and the
measured wavelength-differential visibilities $\delta V(\lambda)$ before a
Levenberg-Marquardt least-square fitting algorithm is used to determine the best-fit model parameters.
The measured closure phases are fit simultaneously with the differential visibilities,
weighting each data point according to the determined error bars (see equations 8--10 in \citealt{kra05}).
Parameters in our binary star model are the binary separation $\rho$, the position angle 
$\Theta$, the intensity ratio $F_{\rm C2}/F_{\rm C1}$, and the uniform
disk diameter of the components.
Since our earlier speckle measurements indicated that $F_{\rm C2}/F_{\rm C1}$
is practically constant in the NIR wavelength range \citep[see ][]{kra07}, we 
assume that the intensity ratio does not vary over the $H$- and $K$-bands.
For the model fitting, we fix the apparent diameters of the two stars to 
$0.22$\,mas for the primary star (corresponding to $10.6~R_{\sun}$ at a distance 
of 450~pc) and $0.15$\,mas for the companion ($7.2~R_{\sun}$, 
using the spectral type determined by \citealt{kra07} and the stellar evolutionary models
of \citealt{mar05}).
Our AMBER LR measurements and best-fit models are shown in Fig.~\ref{fig:fitP78OT} 
and \ref{fig:fitP80Dec}. 
The derived astrometric data are given in Tab.~\ref{tab:astrometry}.

\section{VLTI/AMBER aperture synthesis imaging} \label{sec:imaging}

\begin{figure*}[t]
  \centering
  \includegraphics[width=18cm]{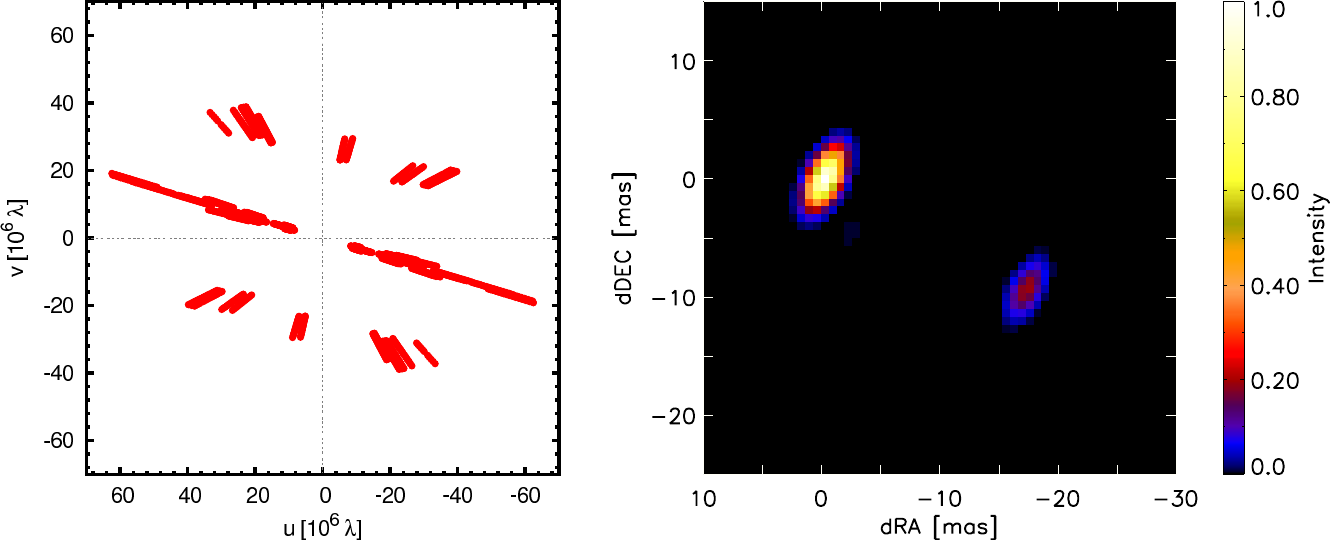}\\
  \caption{
    Combining AMBER data obtained on three telescope configurations ({\it Left:} $uv$-coverage), we reconstructed
    an aperture synthesis image of the {\toric} system with an effective resolution of $\sim 2$\,mas ({\it Right}).
    For a detailed description, we refer to Sect.~\ref{sec:imaging}.
  }
  \label{fig:imaging}
\end{figure*}

Currently, the most commonly applied procedure to extract scientific information 
from optical interferometric data is model fitting, which requires the selection
of an appropriate (geometrical or astrophysically motivated) model,
whose parameters are then adjusted to fit the interferometric observables.
Since this model fitting approach requires {\it a priori} knowledge 
about the source structure, it might, in some cases, not be applicable 
or might lead to biased results.
Therefore, it is highly desirable that optical interferometers
such as VLTI/AMBER have the capability of recovering the source brightness 
distribution free of any assumptions.
The aim of this section is to apply state-of-the-art aperture synthesis 
imaging techniques for the first time to real VLTI/AMBER data 
in order to independently confirm the scientific results obtained for {\toric} 
in the last section, and, simultaneously, to demonstrate 
the imaging capabilities of VLTI/AMBER on a relatively well-studied astrophysical
target with limited intrinsic complexity.

In order to obtain the $uv$-coverage required for aperture synthesis imaging, 
we combined the {\toric} AMBER data sets taken between December 2007 and March 2008
on three different 3-telescope array configurations and at several hour angles.
Assuming that the source morphology does not change significantly over the 
$K$-band (which seems well justified based on our earlier 
measurements of the wavelength-dependent binary flux ratio; \citealt{kra07}), 
we make use of AMBER's spectral coverage, yielding radial 
tracks in the $uv$-plane (Fig.~\ref{fig:imaging}, {\it left}).
Since the $K$-band visibilities provide a more reliable absolute 
calibration (see Sect.~\ref{sec:obsamber}), we did not use the 
$H$-band data for image reconstruction and rejected also one
measurement taken under particularly poor and variable conditions 
(2007~Dec~05, UT~07:46).
In order to correct for the binary orbital motion
over the 3-month period ($\Delta\rho=2.2$\,mas, $\Delta\Theta=6.6$\degr), 
we apply a rotation-compensating coordinate
transformation of the $uv$-plane (see \citealt{kra05} for a description
of this procedure) using the astrometric 
data given in Tab.~\ref{tab:astrometry}.
Then, we employed our 
{\it Building Block Mapping} software, which is based on the 
algorithm described by \citet{hof93}.
Starting from an initial single $\delta$-function, this algorithm 
adds components to a model image in order to minimize
the deviations between the measured bispectrum and the bispectrum of the
model image.
Finally, the image is convolved with a clean beam of 
$\sim 1.5 \times 3$\,mas, reflecting the elongation of the sampled $uv$-plane.
The resulting $K$-band aperture synthesis image 
(Fig.~\ref{fig:image} and Fig.~\ref{fig:imaging}, {\it right}) 
yields a direct, model-independent representation of our 
VLTI/AMBER interferometric data.
Measuring the companion position in the reconstructed image
($\rho=19.3$\,mas, $\Theta=241$\degr, epoch 2007.9), we find good agreement with the results obtained with 
our model fitting approach using wavelength-differential observables (Sect.~\ref{sec:modelfit}).
The noise features within the image show an intensity amplitude 
below 2\% of the peak brightness in the image.

\section{Results}

\subsection{Dynamical orbit of the {\toric} binary system} \label{sec:orbitsolution}

\begin{figure}[tbp]
  \centering
  \includegraphics[width=8.8cm]{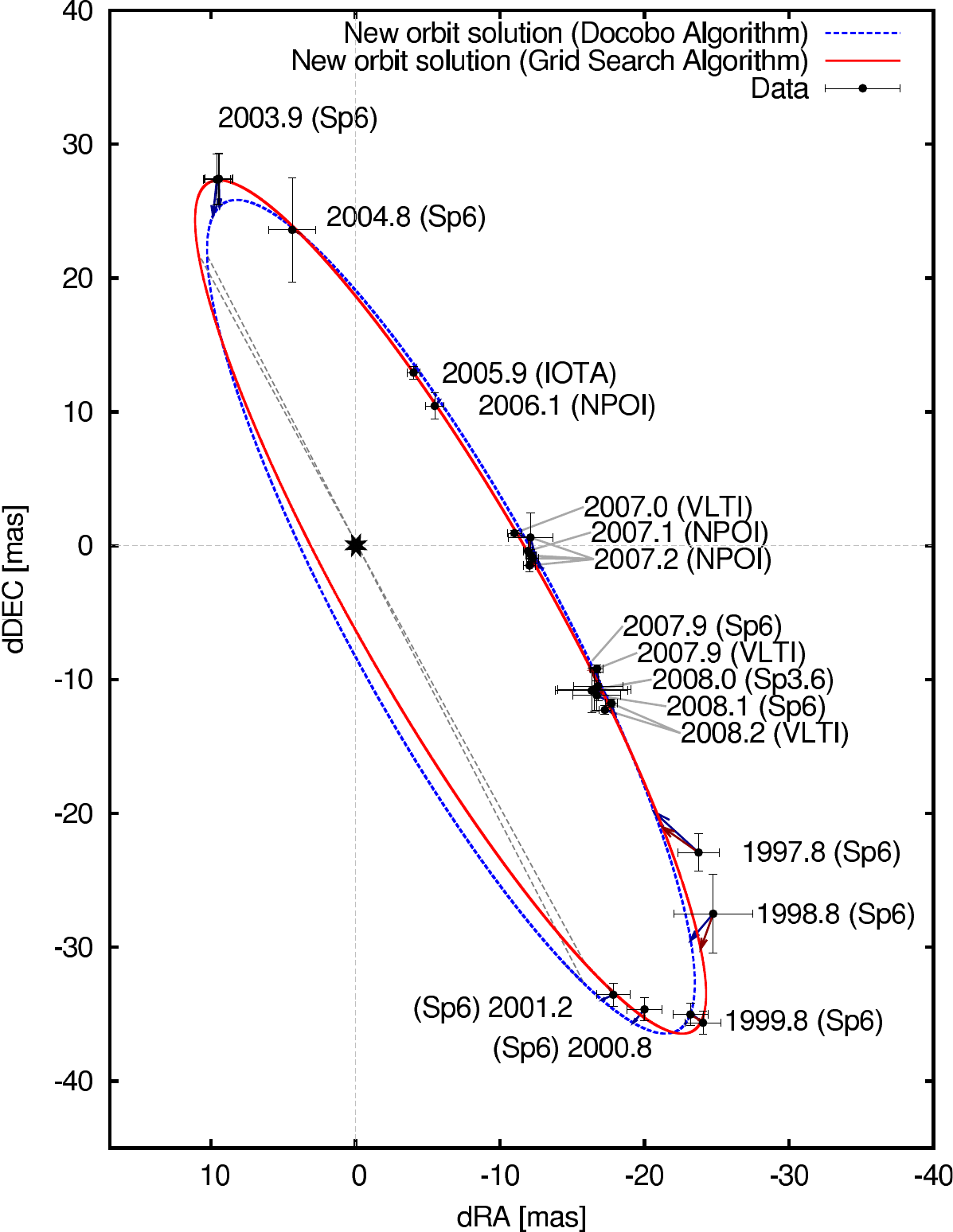} 
  \caption{Comparison of our new orbit solutions 
    with the available astrometric data.
    Each position measurement is connected to the orbit prediction
    with an $O$--$C$ vector (arrows).
    Furthermore, we show the lines of nodes in grey.
    The plots are centered on the primary star.  North is up and east is to the left.
    \vspace{5mm}
  }
  \label{fig:orbit}
\end{figure}

\begin{table}[t]
  \caption{Orbital elements as determined with the algorithm from 
    \citet{doc85} and our grid search algorithm (described in
    Sect.~\ref{sec:orbitsolution}).
  }
\label{tab:orbitalelements}
\centering
\begin{tabular}{ll|cc}
  \hline\hline
                         &         & Docobo          & Grid search \\
  Parameter              &         & algorithm       & algorithm \\
  \hline
  $P$                    & [yrs]   & $11.05 \pm 0.03$      & $11.26 \pm 0.5$    \\
  $T_{0}$                &         & $2002.87 \pm 0.40$    & $2002.57 \pm 0.5$ \\
  $e$                    &         & $0.534 \pm 0.050$     & $0.592 \pm 0.07$   \\
  $a$                    & [mas]   & $40.00 \pm 3.00$      & $43.61 \pm 3$    \\
  $i$                    & [\degr] & $100.7 \pm 1.0$       & $99.0 \pm 2.6$    \\
  $\Omega$               & [\degr] & $25.3 \pm 1.5$        & $26.5 \pm 1.7$     \\
  $\omega$               & [\degr] & $290.9 \pm 2.5$       & $285.8 \pm 8.5$     \\
  $\chi^2_r$              &         & 1.84                 & 0.56 \\
  \hline
  $a^{3}/P^{2}$           & [mas$^3$/yrs$^2$] & $524 \pm 130$    & $645 \pm 200$ \\
  $M_{\rm C1}/M_{\rm C2}$  &                   & $0.21 \pm 0.05$  & $0.23 \pm 0.05$ \\
  $M_{\rm C1}+M_{\rm C2}$  & [$M_{\sun}$]       & $49 \pm 4$       & $47 \pm 4$  \\
  $d_{\rm dyn}$           & [pc]              & $456 \pm 13$     & $416 \pm 12$   \\
  \hline
\end{tabular}
\begin{minipage}{\linewidth}
\vspace{2mm}
{\it Notes}~--~Besides the orbital elements, we give the 
    mass ratio (Sect.~\ref{sec:massratio}), 
    dynamical distance, and system mass (Sect.~\ref{sec:dynmassparallax}),
    derived from both set of orbit elements.
    The dynamical distance and system mass was determined using the
    method from \citet[][method~{\it c} in Sect.~\ref{sec:dynmassparallax}]{bai46} and three different MLRs.
    When assuming another distance $d^{\prime}$, the dynamical 
    system mass $M_{\rm C1}+M_{\rm C2}$ must be scaled by a factor $(d^{\prime}/d_{\rm dyn})^3$.
    The mass ratio $M_{\rm C1}/M_{\rm C2}$ was also computed for the distance $d_{\rm dyn}$,
    but can be converted to any other distance using equation~\ref{eqn:massratio}.
\end{minipage}
\end{table}

\begin{figure*}[tp]
  \centering
  \includegraphics[width=18cm]{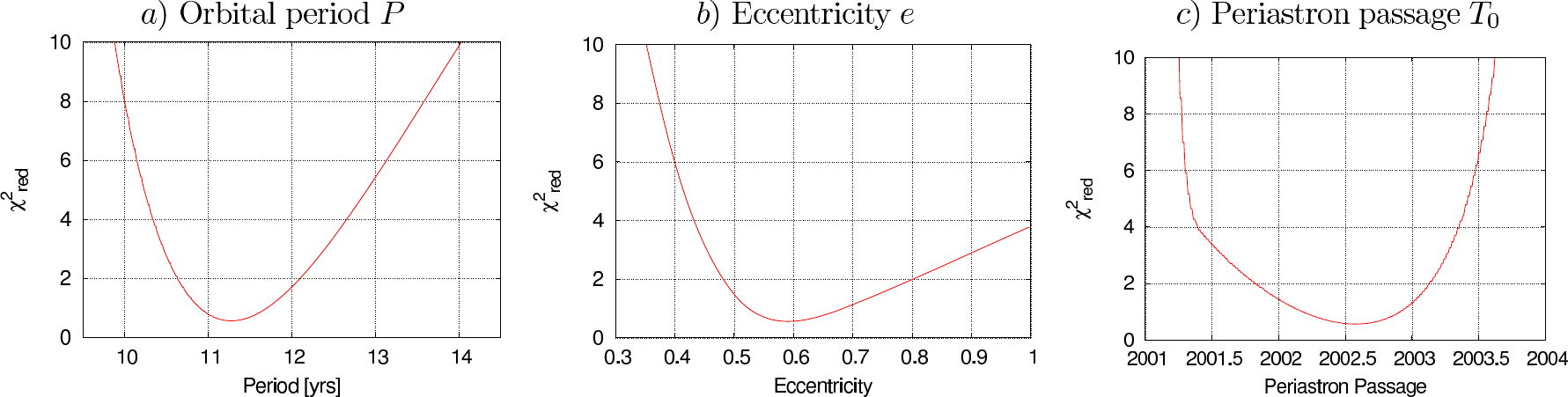}\\
  \caption{Minimum $\chi^{2}_{\rm r}$ curves as function of the 
    dynamical orbital parameters $P$, $e$, $T_{0}$.
    These curves were used to determine the best-fit orbit solution
    and to evaluate the uncertainties on the individual parameters
    (see Tab.~\ref{tab:orbitalelements}).
  }
  \label{fig:orbitchisq}
\end{figure*}

Since our new astrometric data extend the orbital coverage for the
{\toric} system by about 12~months and 
solves the calibration problem described in Sect.~\ref{sec:obsspeckle}, 
we can significantly improve the orbital solution of the system.
To derive a refined orbital solution, we applied two independent orbit fitting 
approaches.

First, we used the method presented by \citet{doc85},
which generates a family of Keplerian orbits, whose 
apparent orbits pass through three base points.
These base points might be selected from the actual
astrometric measurements or represent observationally
favored areas in the $(\rho, \Theta, t)$ parameter space.
From the generated class of possible solutions, the orbit  
which best agrees with the measured separations and PAs is selected.  We use the error
bars of the individual measurements as weight.  
The orbital elements for the determined best-fit orbit solution are given in 
Tab.~\ref{tab:orbitalelements}.

In addition, we implemented a grid search algorithm which scans the parameter space of the
dynamical elements eccentricity $e$, period $P$, and time of periastron passage $T_{0}$.
As described by \citet{hil01}, at each grid point, the geometrical orbital elements $a$, $i$,
$\Omega$, and $\omega$ can be determined by fitting the Thiele-Innes constants to the
observational data.
We scanned the parameter space between 
$P=9 \ldots 16$~yrs (in increments of 0.001~yrs), 
$e=0.0 \ldots 1.0$ (in increments of 0.005), and 
$T_{0} = 2000.0 \ldots 2010.0$ (in increments of 0.001~yrs) and determined 
the least-square distance between the $N$ measured positions $(\rho_{i}, \Theta_{i})$ and 
the corresponding orbit positions $(\rho^{\prime}_{i}, \Theta^{\prime}_{i})$ for each
orbit solution:
\begin{equation}
  \chi^{2}_{\rm r} = \frac{1}{N} \sum_{i=1 \ldots N} \left[ \left( \frac{\rho_{i} - \rho^{\prime}_{i}}{\sigma_{\rho_{i}}} \right)^{2} + \left( \frac{\Theta_{i} - \Theta^{\prime}_{i}}{\sigma_{\Theta_{i}}} \right)^{2} \right]
\end{equation}
Using the $\chi^{2}_{\rm r}$ value determined at each grid point, we built a $\chi^{2}_{\rm r}(P, e, T_{0})$ 
data cube, which we projected to determine the local minimum and 
the associated uncertainty for each parameter.
Fig.~\ref{fig:orbitchisq} shows the determined $\chi^{2}_{\rm r}$ minima curves, from which we
determine the best-fit orbital elements given in Tab.~\ref{tab:orbitalelements}.
Given that the orbital coverage has been substantially improved since our 2007 study,
we do not have to impose {\it a priori} constraints on the distance/system mass in 
order to eliminate unphysical orbit solutions.
In Fig.~\ref{fig:orbit}{\it c} we compare the orbital solutions determined with 
the \citet{doc85} algorithm and with the grid search algorithm 
to the available astrometric data.

\subsection{Constraining the binary mass ratio} \label{sec:massratio}

\begin{figure}
  \centering
  \includegraphics[width=8.5cm]{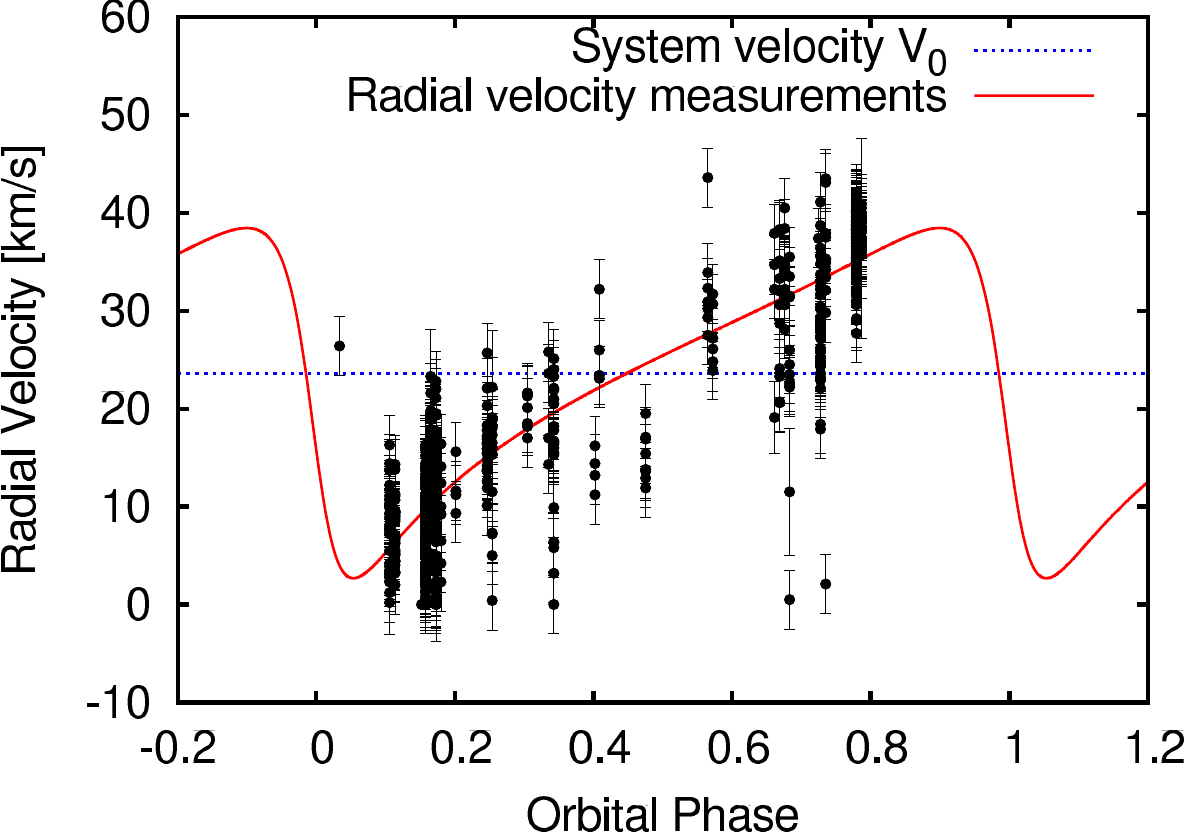} \\
  
  \caption{The radial velocities measured towards {\toric} plotted as function of orbital phase (assuming the 
    orbital elements determined with our grid search algorithm; see Tab.~\ref{tab:orbitalelements}).  
    Besides the extensive data set by \citet{sta08}, we included the
    radial velocities measured by \citet{str44}, \citet{con72}, and \citet{mor91}.
    The red line shows the radial velocities corresponding to our orbital solution with a system velocity 
    $V_{0} = 23.6$~km\,s$^{-1}$ (blue dashed line) and the mass ratio $q(414~\mathrm{pc})=0.23$
    as determined from our least-square fit (Sect.~\ref{sec:massratio}).
  }
  \label{fig:radvel}
\end{figure}

\begin{figure}
  \centering
  \includegraphics[width=8.5cm]{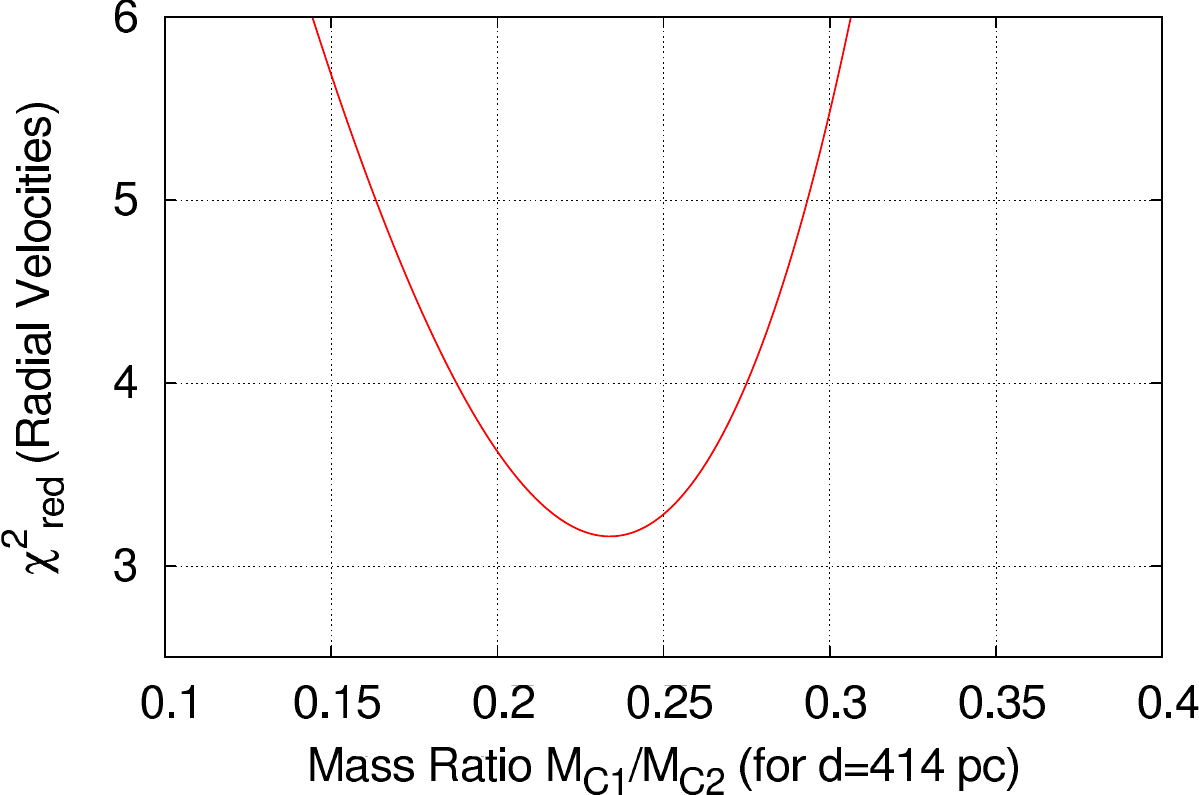} \\
  
  \caption{$\chi^{2}_{\rm r}$ curve for our fit of the binary mass ratio to the 
    available radial velocity data (assuming a distance of $414$~pc).
    We assume the orbital elements (Tab.~\ref{tab:orbitalelements}) derived from our 
    interferometric measurements.
  }
  \label{fig:radvelchisq}
\end{figure}

Besides the $15.424$-day period which is associated with the wind from the primary, long-term radial 
velocity variations were also found \citep{vit02,sta08}.
Using a large data set covering more than 15~yrs of spectroscopic observations (plus 
three archival measurements, which extend the coverage to more than 64~yrs), \citet{sta08} showed 
that these variations are consistent with the orbital motion of a high-eccentricity binary system.
Although the strong scatter within the radial velocity measurements prevents us from solving for
the precise spectroscopic orbit, the combination of these data with our new orbital solution
can be used to provide a first direct constraint on the mass ratio of the components in 
the {\toric} system.

In Fig.~\ref{fig:radvel} we plot the available radial velocity data as a function of orbital 
phase using the orbital period $P$ and time for periastron passage $T_{0}$ determined 
independently from our interferometric measurements (Sect.~\ref{sec:orbitsolution}).
Using the method from \citet{pou98}, we compute the radial velocity variations corresponding to our
full set of orbital elements (Tab.~\ref{tab:orbitalelements}) and perform a 
least-square fit between the measured ($v_{i}$) and the predicted ($v^{\prime}_{\rm i}$) radial velocities 
in order to determine the mass-ratio between the binary components. 
As least-square measure for $N$ measurements, we use
\begin{equation}
  \chi^{2}_{\rm r} = \frac{1}{N} \sum_{i=1 \ldots N} \left( \frac{v_{i} - v^{\prime}_{i} - V_{0}}{\sigma_{v_{i}}} \right)^{2},
  \label{eqn:chisqradvel}
\end{equation}
where $V_{0}$ is the velocity of the center of mass of the system, which is given by 
$V_{0} := \langle v_{i} \rangle - \langle v^{\prime}_{i} \rangle$.
Since the radial velocities were extracted from various spectral lines (\ion{C}{iv}, \ion{He}{ii}, and 
\ion{O}{iii}) and it is known that these lines can show, with respect to each other, systematic velocity offsets 
on the order of $2-3$~km\,s$^{-1}$ \citep{sta08}, we used 3~km\,s$^{-1}$ as minimum velocity error 
$\sigma_{v_{i}}$ in order to avoid overweighting individual measurements.
By varying the mass-ratio between the components, we find 
\begin{eqnarray}
  q(d) & := & \frac{M_{\rm C1}}{M_{\rm C2}} = \left( 5.27^{+1.17}_{-0.75} \frac{d}{414~\mathrm{pc}} -1 \right)^{-1},
  \label{eqn:massratio}
\end{eqnarray}
i.e.\ $q(450~\mathrm{pc}) = 0.21 \pm 0.04$ or $q(414~\mathrm{pc}) = 0.23 \pm 0.05$,
which is slightly lower than the 
value we derived earlier by modeling the wavelength-dependent binary flux ratio of the {\toric} 
system \citep[$q = 0.45 \pm 0.15$,][]{kra07}.

For the radial velocity of the center of mass, we determine $23.6$~km\,s$^{-1}$, 
which is in good agreement with the heliocentric velocity of the 
Orion Molecular Cloud \citep[$\sim 28$~km\,s$^{-1}$, ][]{ode01}.
This might indicate that the relative motion of the {\toric} system 
with respect to the parental cloud is smaller than previously assumed
\citep{ode01,sta08}.

\subsection{Dynamical masses and parallaxes} \label{sec:dynmassparallax}

\begin{figure}
  \centering
  \includegraphics[width=8.5cm]{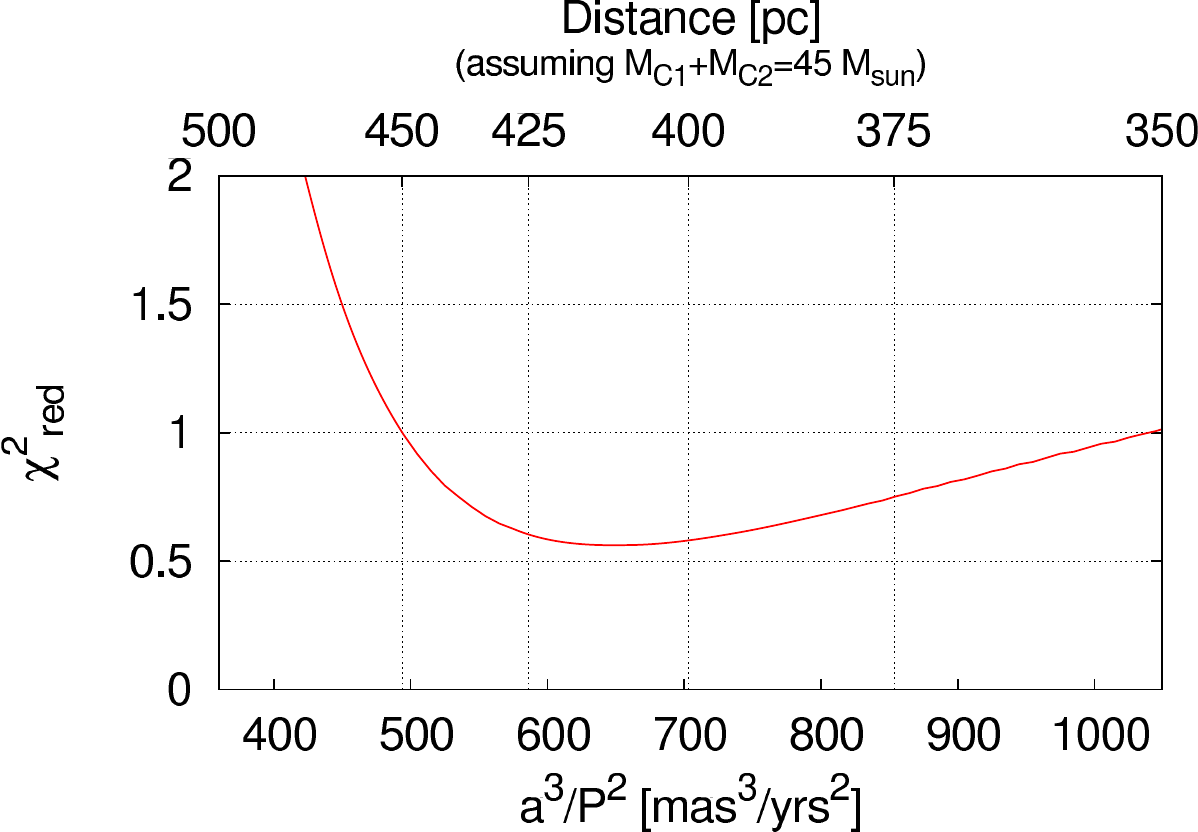} \\
  
  \caption{Minimum $\chi^{2}_{\rm r}$ curve as a function of 
    $a^{3}/P^{2} = (M_{\rm system}) \cdot \pi^{3}$, through
    which the dynamical orbital parameters can be related to
    the mass sum and the dynamical parallax.
    For illustration, we give the dynamical distance 
    in the upper axis assuming a total system mass of 45~$M_{\sun}$.
  }
  \label{fig:a3p2chisq}
\end{figure}

\begin{figure}
  \centering
  \includegraphics[width=8.5cm]{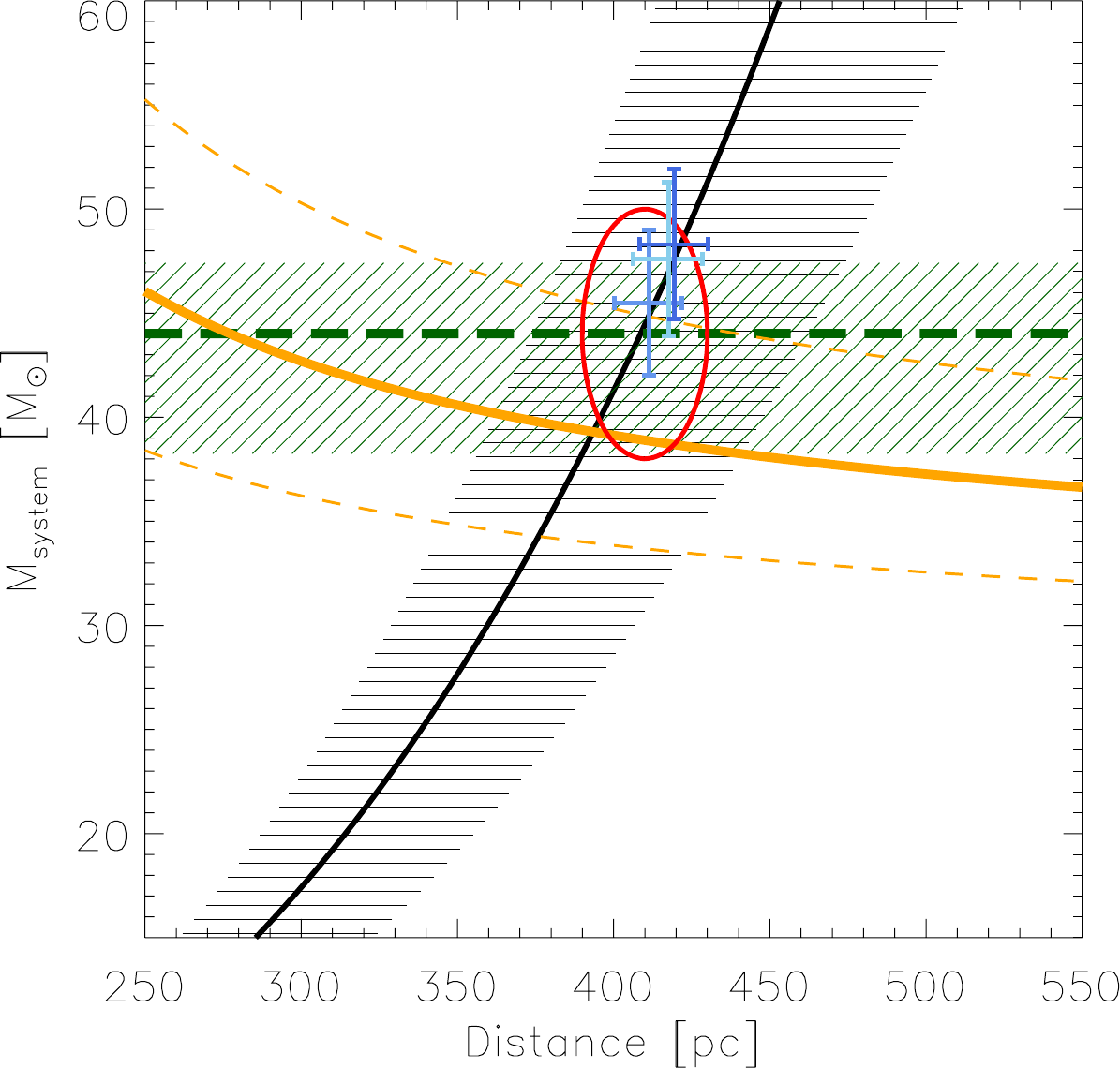} \\
  
  \caption{
    The derived $a^3/P^2$ value puts some direct constraints 
    on the system mass $M_{\rm system} := M_{\rm C1}+M_{\rm C2}$ and the dynamical distance.
    As described in Sect.~\ref{sec:dynmassparallax}, we employ three 
    alternative methods to obtain constraints, which help to disentangle 
    the system mass and the distance 
    (method {\it a:} dashed green-shaded area,
    method {\it b:} orange curve, 
    method {\it c:} blue data points).
    The black-shaded area marks the $a^3/P^2$-constraints 
      determined with our orbit grid search algorithm.
    With the red ellipse, we visually mark the area of best agreement between all constraints.
  }
  \label{fig:msumdistchisq}
\end{figure}

Kepler's third law ($M_{\rm system} \cdot \pi^3 = a^3/P^2$)
relates $a$ and $P$ with the product of the system mass $M_{\rm system} := M_{\rm C1}+M_{\rm C2}$
and the cube of the parallax $\pi$.
Therefore, our astrometric measurement of $a^3/P^2$ directly constrains 
certain areas in the ($\pi$,~$M_{\rm system}$) parameter space, 
as shown by the shaded area in Fig.~\ref{fig:msumdistchisq}.
Since the elements of the spectroscopic orbit are currently only 
weakly constrained, it is not possible to directly separate the 
system mass and the dynamical parallax.
However, several independent methods exist to disentangle these important
parameters using additional information, such as the flux ratio 
of the components, their mass ratio, and/or their stellar parameters.

The stellar parameters of {\toricone} have already been studied extensively
with spectroscopy, placing the effective temperature between 37\,000 and 40\,000~K 
\citep[e.g.][]{rub91,bal91,sim06,pol07}.  In particular, \citet{sim06}
included non-LTE and line-blanketing effects ($T_{\rm eff,C1}=39\,000 \pm 1000$~K)
for their detailed spectroscopic modeling, 
which makes their results compatible with the O-star stellar parameter 
calibration by \citet{mar05}.

Since each of the above-mentioned stellar or observational parameters 
is associated with certain assumptions and uncertainties, it seems advisable to 
take a number of alternative approaches for deriving the underlying
physical parameters in order to yield some insight into the
associated uncertainties.
Therefore, we follow three alternative approaches:
\begin{enumerate}
\item[\it a)] Based on the effective temperature determination of the 
  primary and the binary flux ratio, one can give a reasonable 
  mass range for each component and, thus, the system mass
  (green-shaded area in Fig.~\ref{fig:msumdistchisq}).
\item[\it b)] One can estimate the
  mass of the primary from the stellar temperature and then 
  derive the companion mass using the mass ratio constraints
  obtained from the radial velocities (Sect.~\ref{sec:massratio}), 
  yielding the orange curve in Fig.~\ref{fig:msumdistchisq}.
\item[\it c)] \citet{bai46} presented a method which solves for the
  system mass and the dynamical parallax using a 
  mass-luminosity relation (MLR), the bolometric corrections of the
  components and their extinction-corrected apparent magnitudes. 
  To evaluate the influence of the MLRs on the result, we used three
  different MLRs; namely, from \citet{bai46}, \citet{hei78}, and \citet{dem91}
  (corresponding to the three blue data points in Fig.~\ref{fig:msumdistchisq}).
\end{enumerate}

For methods {\it a)} and {\it b)}, we employ the stellar calibration by 
\citet{mar05}, while for method {\it c)}, three earlier calibrations
are used.  
In method {\it a)} and {\it b)}, the stellar temperature of the primary component 
is fixed to the value by \citet[][ $T_{\rm eff,C1}=39\,000 \pm 1000$~K]{sim06},
while in {\it c)} we scan a slightly wider range of temperature values ($T_{\rm eff, C1}=37\,000 \ldots 40\,000$~K, 
$T_{\rm eff, C2}=30\,000 \ldots 33\,000$~K, using the bolometric correction by \citealt{bes98}).
Method {\it a)} requires the $V'$-band flux ratio ($F_{C2}/F_{C1}=0.31 \pm 0.02$)
and {\it c)} the extinction-corrected magnitude of the total system 
\citep[$V=5.12 \pm 0.1$, $A_{V}=1.74 \pm 0.1$,\ ][]{hil97}.

Evidently, each method is associated with considerable uncertainties, which
makes it very desirable for future observations to improve not only the astrometric
orbit, but to derive the accurate spectroscopic orbit of the system as well.
Nevertheless, within their large uncertainties, the methods employed cover a
common area in parameter space, corresponding to a system mass of
$M_{\rm system}=44 \pm 7~M_{\sun}$ and a dynamical distance of
$d=410 \pm 20$~pc, as marked with the red ellipse in Fig.~\ref{fig:msumdistchisq}.
Methods {\it a)} and {\it b)} yield systematically lower system masses than
method {\it c)}, reflecting the recent correction in the mass calibration scale
in stellar evolutionary models \citep{mar05}.
Assuming a distance of 414~pc, as determined by \citet{men07}, 
would yield a system mass of $46~M_{\sun}$.

\subsection{Possible implications on the dynamical history of the {\toric} system}

With an eccentricity of $\sim 0.6$, the orbit of {\toric} is 
located on the upper end of the eccentricity distribution of low-
as well as high-mass binary stars \citep{mat94,mas98}, perhaps providing
important information about the dynamical history of the system.
\citet{tan04,tan08} proposed that the \object{Becklin-Neugebauer} (BN) object, which
is located 45\arcsec\ northwest of the Trapezium stars, might be a
runaway B star ejected from the {\toric} multiple system approximately
$4\,000$~yrs ago.  This scenario is based on proper motion measurements, which
show that BN and {\toric} recoil roughly in opposite directions.
Three-body interaction is a crucial part of this interpretation, and 
the high eccentricity of the {\toric} orbit which we derive for this system
might be a direct consequence of this dynamical interaction event.
However, another study \citep{rod05} also aimed to identify the multiple
system from which BN was ejected and identified \object{Source~I} as the likely
progenitor system.  
Later, \citet{gom05,gom08} added further evidence to this
interpretation by identifying \object{Source~$n$} as a potential third
member of the decayed system.
Therefore, it is still unclear whether this scenario can explain
the measured properties of the {\toric} orbit.

As an alternative explanation for the high eccentricity of the
{\toric} system, \citet{zin07} pointed out that such systems 
are predicted by star formation scenarios which include 
sub-Keplerian rotating disks or filament fragmentation \citep{kra06}.

\section{Conclusions}

We have presented new bispectrum speckle ($V'$-/$B'$-band) and 
VLTI/AMBER ($H$-/$K$-band) interferometric 
observations of the Orion Trapezium star {\toric}
covering several epochs over a time period of about 14~months.

From our long-baseline interferometric data, we have reconstructed the first 
model-independent VLTI/AMBER aperture synthesis image, 
depicting the {\toric} system at a resolution of $\sim 2$\,mas and
demonstrating the imaging capabilities of this unique facility.
In order to extract accurate astrometric information for all epochs, 
we have followed a new modeling approach which is based on 
wavelength-differential observables and which
demonstrates the benefits of spectro-interferometry in terms 
of observing efficiency and robustness to poor observing conditions.
Furthermore, our ESO~3.6\,m and the BTA~6\,m speckle observations
allow us to solve 180\degr-ambiguity and to calibrate the 
closure phase sign of our VLTI/AMBER observations, providing a 
potential reference for other AMBER observations using
closure phase information~\footnote{Since our data set might be useful for VLTI/AMBER users as reference
data for the calibration of the closure phase sign, we provide our data on the following website:
{\tt http://www.mpifr.de/staff/skraus/files/amber.htm}}.

Our new {\toric} astrometric data shows that since its discovery in 1997, 
the {\toric} companion has nearly completed one orbital revolution and that the 
system has a high eccentricity ($e \sim 0.6$).
Solving for the orbital elements,
we determine a period of $\sim$11.3~yrs, 
a semi-major axis of 44\,mas, and a periastron passage around 2002.6.
According to our orbital solutions, the physical separation between the
components decreases to $\sim 7$~AU (2.8\,mas) during periastron passage, 
which might be too large to result in detectable signatures of wind-wind interaction 
between the stellar winds.

Using additional information about the stellar parameters 
and various stellar models, we estimate the total system 
mass to be $44 \pm 7~M_{\sun}$ and the dynamical distance to be $d=410 \pm 20$~pc.
In the coming years, the uncertainties on these parameters could be considerably 
reduced with new spectroscopic and astrometric observations of this 
important binary system.

\begin{acknowledgements}
We would like to thank the referee, M.~McCaughrean, for helpful comments which improved this paper.
Furthermore, we acknowledge helpful discussions with E.~Vitrichenko, 
F.~Millour and members of the AMBER consortium.
\end{acknowledgements}

\begin{appendix}

\section{Spectral calibration of AMBER-LR data} \label{app:speccalib}

\begin{figure*}[tp]
  \centering
  \includegraphics[width=18cm]{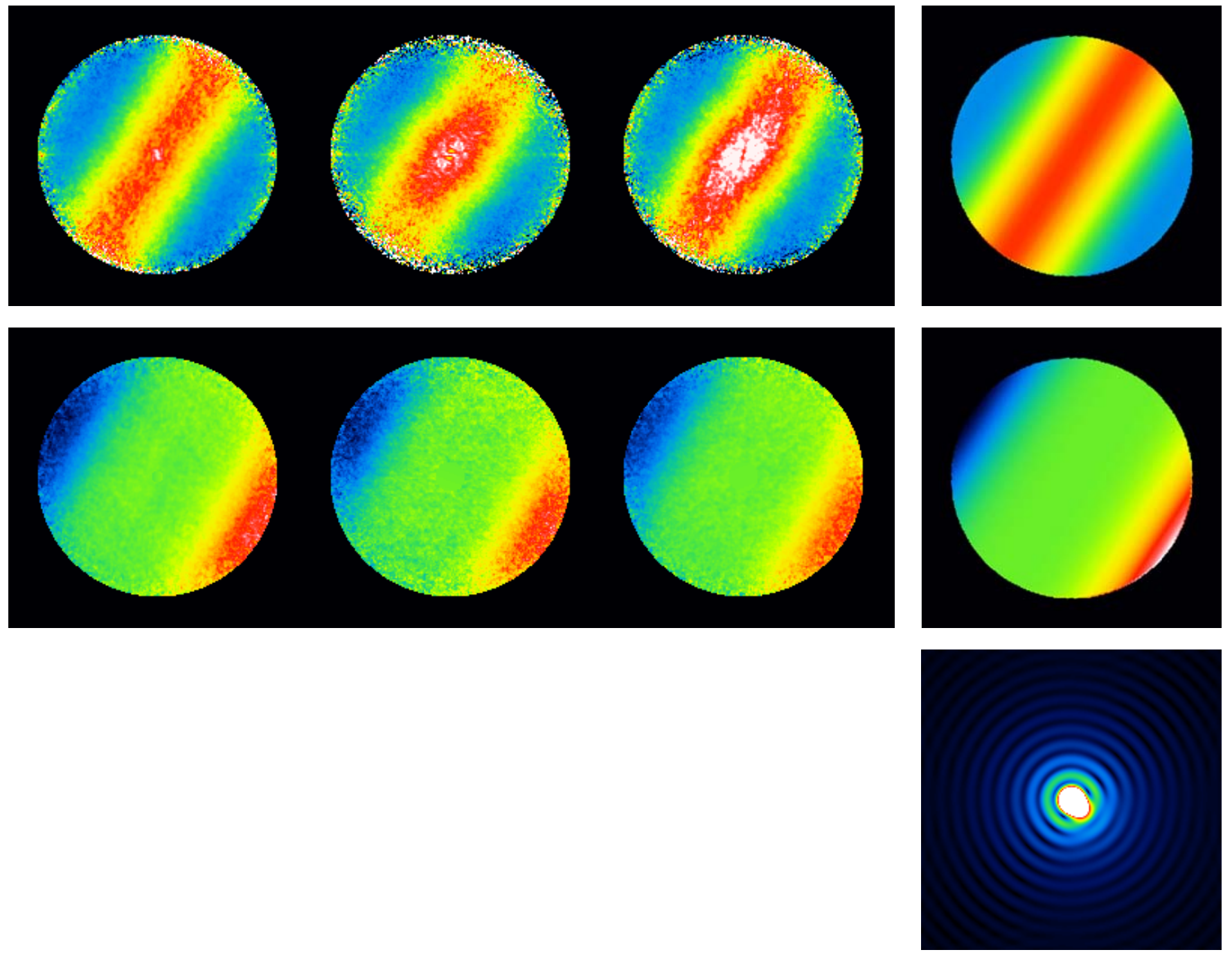} \\
  \caption{
    {\it Left:} Three two-dimensional average power spectra ({\it top row}) and 
    Fourier phase spectra ({\it middle row}) derived 
    from {\it B'}-band speckle data recorded on 2008~Jan~10 with the ESO~3.6\,m telescope. 
    {\it Right:} Comparing this data to model power spectra ({\it top row}) and Fourier spectra ({\it middle row})
    for a binary with separation $\rho=19.7$\,mas and PA {$\Theta=237$\deg} shows that the 
    fainter component is located southwest of the primary star (the bispectrum speckle image 
    reconstructed from the model visibilities and phases are shown to the bottom right).
    Together with the calibration data sets mentioned in Sect.~\ref{sec:obsspeckle}, 
    this data set unambigously defines the current orientation of the {\toric}
    binary system and shows that the system has nearly completed one orbital revolution
    since its discovery in 1997.
  }
  \label{fig:speckle}
\end{figure*}

\begin{figure}[tbp]
  \centering
  \includegraphics[width=8.5cm,angle=0]{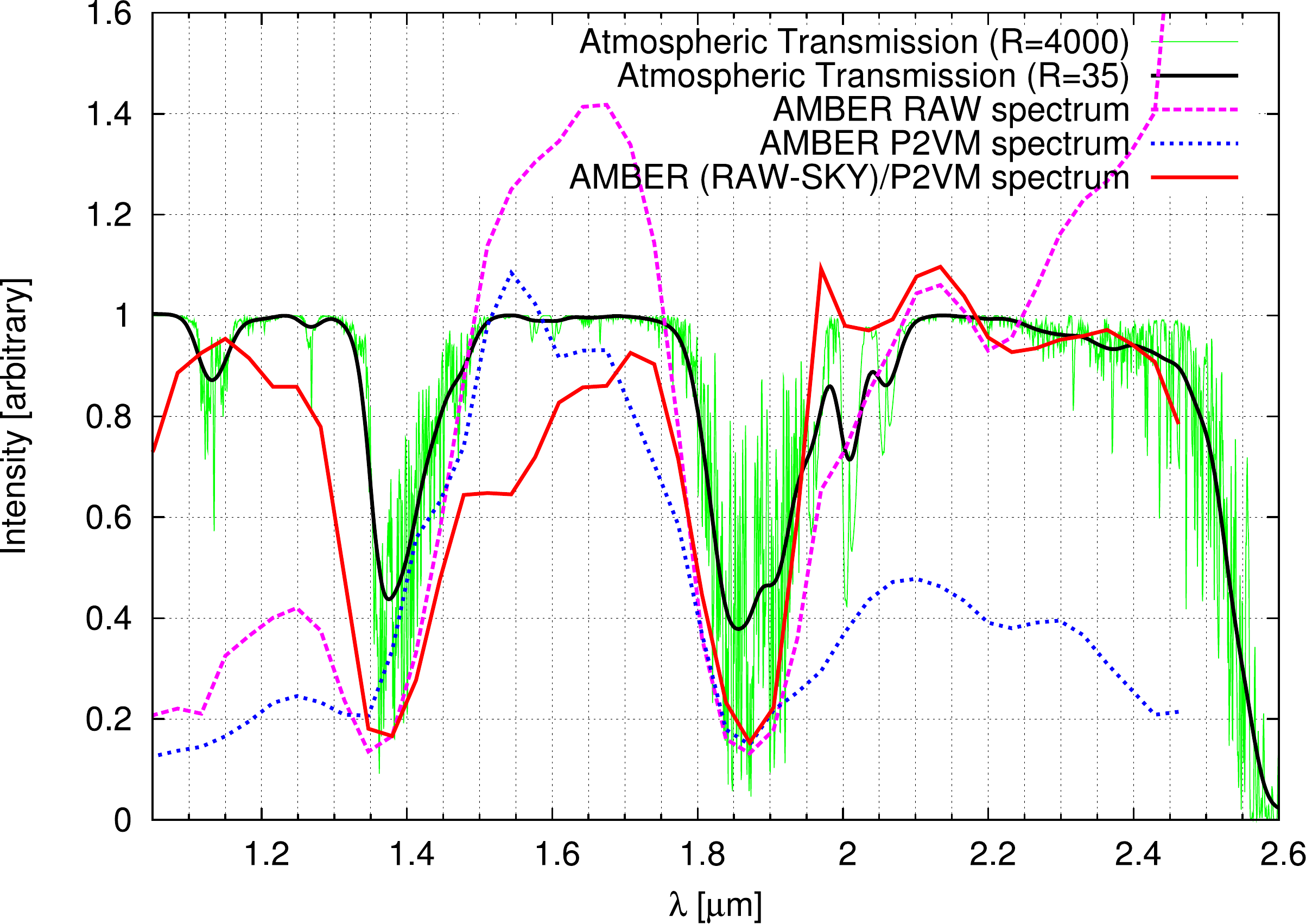}\\
  \caption{
    To obtain a spectrum which is corrected for instrumental effects for the spectral calibration, 
    we subtract from the RAW spectrum the SKY spectrum and then divide by the P2VM spectrum (blue dashed curve).
    The resulting instrument-corrected spectrum (red curve) is compared with an 
    atmospheric transmission spectrum (green curve: $R=4000$; black curve: $R=35$).
  }
  \label{fig:speccalib}
\end{figure}

Employing wavelength-differential visibilities and phases for model fitting
requires a precise knowledge of the central wavelength of the recorded spectral channels.
Therefore, we performed a re-calibration of the wavelength axes of our data sets
using atmospheric absorption features and found deviations of up to $\sim 0.1~\mu$m
compared to the standard wavelength table applied to the data on Paranal.

In the recorded $J$-/$H$-/$K$-band object raw 
spectra, the telluric features are not pronounced enough to be used for the calibration, 
which is mainly due to absorption by internal optical components.
Therefore, besides the raw spectrum of the object exposures, we also extracted the
spectrum from the sky and the P2VM calibration exposures.  
The P2VM calibration files (P2VM={\it Pixel-to-Visibility-Matrix}; see \citealt{tat07b}) 
are recorded at the beginning of each observation block
using a lamp located in AMBER's {\it Calibration and Alignment Unit} and 
provide a measure of the instrumental transmission.
By subtracting the sky spectrum from the object spectrum and then 
dividing by the P2VM spectrum, we yield a spectrum which is corrected for
most instrumental effects and shows the telluric absorption features much more clearly
(see Fig.~\ref{fig:speccalib}, red curve).
These corrected spectra were compared with standard atmospheric transmission spectra
provided by the Gemini observatory\footnote{These ATRAN transmission spectra can be found on the website 
{\tt http://www.gemini.edu/sciops/telescopes-and-\newline sites/observing-condition-constraints/\newline transmission-spectra}.}
(see Fig.~\ref{fig:speccalib}, green curve).
We manually align the spectra using the pronounced gaps between the $J$-/$H$- and $H$-/$K$-band and 
find best agreement assuming a linear dispersion law $\lambda(i) \propto 0.0328 \cdot i$, where
$i$ is the number of the spectral channel on the detector.

The remaining wavelength calibration uncertainty of about 1 spectral channel
(corresponding to about $0.03~\mu$m) is the dominant limiting factor on the achievable 
astrometric accuracy ($\sim 2$\%).
Therefore, for future observations, it seems highly desirable to implement an
AMBER on-site spectral calibration device (e.g.\ NIR lasers with well-known frequencies),
fascilitating an absolute spectral calibration in the standard instrument calibration
procedure, pushing AMBER to its full astrometric accuracy.
\end{appendix}

\bibliographystyle{aa}
\bibliography{10368}

\end{document}